\documentclass[12pt]{article}

\usepackage{subfig}
\usepackage{color}
\usepackage{epsfig,amssymb,amsfonts,amsmath,graphicx,dsfont,cite,xfrac}
\usepackage{authblk}
\usepackage{rotating}

\usepackage{tikz,pgfplots}
\usetikzlibrary{decorations.pathreplacing}
\usetikzlibrary{positioning}
\usetikzlibrary{matrix}

\usepackage{cite}
\usepackage[colorlinks=true,linkcolor=blue,citecolor=red]{hyperref}

\newcommand{\be}{\begin{equation}}
\newcommand{\ee}{\end{equation}}
\newcommand{\bea}{\begin{eqnarray}}
\newcommand{\eea}{\end{eqnarray}}

\title{Fourth Painlev\'e and Ermakov equations: quantum invariants and new exactly-solvable time-dependent Hamiltonians}

\author[${1}$]{K. Zelaya}
\author[${2}$]{I. Marquette}
\author[${1,3}$]{V. Hussin}

\affil[${1}$]{\footnotesize Centre de Recherches Math\'ematiques, Universit\'e de Montr\'eal, Montr\'eal H3C 3J7, QC, Canada}

\affil[${2}$]{\footnotesize School of Mathematics and Physics, The University of Queensland, Brisbane, QLD 4072, Australia}

\affil[${3}$]{\footnotesize D\'epartement de Math\'ematiques et de Statistique, Universit\'e de Montr\'eal, Montr\'eal H3C 3J7, QC, Canada}

\date{}

\begin{document}

\maketitle

\begin{abstract}
In this work, we introduce a new realization of exactly-solvable time-dependent Hamiltonians based on the solutions of the fourth Painlev\'e and the Ermakov equations. The latter is achieved by introducing a shape-invariant condition between an unknown quantum invariant and a set of third-order intertwining operators with time-dependent coefficients. The new quantum invariant is constructed by adding a deformation term to the well-known parametric oscillator invariant. Such a deformation depends explicitly on time through the solutions of the Ermakov equation, which ensures the regularity of the new time-dependent potential of the Hamiltonian at each time. On the other hand, with the aid of the proper reparametrization, the fourth Painlev\'e equation appears, the parameters of which dictate the spectral behavior of the quantum invariant. In particular, the eigenfunctions of the third-order ladder operators lead to several sequences of solutions to the Schr\"odinger equation, determined in terms of the solutions of a Riccati equation, Okamoto polynomials, or nonlinear bound states of the derivative nonlinear Schr\"odinger equation. Remarkably, it is noticed that the solutions in terms of the nonlinear bound states lead to a quantum invariant with equidistant eigenvalues, which contains both an (N+1)-dimensional and an infinite sequence of eigenfunctions. The resulting family of time-dependent Hamiltonians is such that, to the authors' knowledge, have been unnoticed in the literature of stationary and nonstationary systems.
\end{abstract}


\section{Introduction}
During the last several decades, physicists have realized the importance of nonlinear equations in the study of physical systems, even in cases where linear equations govern the dynamical laws. For instance, in quantum mechanics, the nonlinear Riccati equation~\cite{Inc56,Sch18} has played a fundamental role in the construction and study of new exactly-solvable models, for it relates the factorization method~\cite{Sch40a,Sch40b,Sch41,Inf51,Mie84,Mie04}, with  the Darboux transformation~\cite{Mat91}. The latter can be formulated in the so-called \textit{supersymmetric quantum mechanics} (SUSYQM)~\cite{Kha93,Car00,Coo01,Don07}. In this regard, systems with spectrum on-demand are obtained either by adding energy level not present in the original model, or by removing levels through the Darboux-Crum transformation~\cite{Cru55}. This formalism represents an outstanding progress in the study of quantum systems; it has allowed extending the families of exactly-solvable models~~\cite{Mie00,Fer99} beyond the conventional harmonic oscillator, hydrogen atom, the interaction between diatomic molecules, and some few others. Moreover, quantum mechanics in the non-Hermitian regime has been explored by generalizing the factorization method and allowing the Riccati equation to be a complex-valued function. In this form, the realitiy of the spectrum is preserved~\cite{Ros15,Ros18,Bla18} in systems with either broken and unbroken $PT$-symmetry, extending the conventional systems with $PT$-symmetry and real spectrum~\cite{Zno00,Bag01,Cor15}. From the latter, the nonlinear Ermakov equation~\cite{Erm08,Pin50,Mil30} emerges naturally from the complexified Riccati one, the solutions of which ensures the regularity of the new complex-valued potentials. For more details on the applications of the Riccati and Ermakov equations in physics, see~\cite{Sch18}.

Among other nonlinear equations, we have the Painlev\'e transcendentals, a family of six nonlinear equations $P_{\rm I}$-$P_{\rm VI}$ with complex parameters, whose solutions are in general transcendental, that is, theyr are not be expressed in terms of classical functions~\cite{Gro02,Mar06}. Nevertheless, for some specific values of the parameters, a seed function can be used to generate a complete hierarchy of solutions through the B\"acklund transformation~\cite{Bas95}, which can be thought as a nonlinear counterpart of the recurrence relations. In particular, the fourth Painlev\'e equation can be taken into a Riccati equation with the appropriate choice of the parameters. we thus solve a ``simpler'' nonlinear equation instead. Also, the fourth Painlev\'e equation has also brought new result in the trend of orthogonal polynomials, where new families were discovered though the hierarchies of rational solutions in terms of the generalized Okamoto, generalized Hermite and Yablonskii-Vorob'ev polynomials~\cite{Cla03}. The Painlev\'e transcendental have also found interesting applications in the study of physical models in nonlinear optics~\cite{Flo90}, quantum gravity~\cite{Fok91}, and SUSYQM~\cite{And00,Mar09,Ber16}, to mention some. Interestingly, the fourth Painlev\'e equation arises quite naturally in third-order shape-invariant SUSYQM~\cite{And00}, where the parameters of the Painlev\'e transcendental define the eigenvalues of the new Hamiltonians. The respective intertwining operators serve at the same time as ladder operators, from where the eigenfunctions are determined. A striking feature of this approach is that, in general, the so-constructed intertwining operators are not in general factorizale in terms of first-order operators. Thus, the results obtained in this way generalize those of~\cite{Suk97}. It is worth to notice that higher-order ladder operators have been also studied, in a different way, in the context of supersymmetric (SUSY) partners for the stationary oscillator in both the Hermitian~\cite{Fer99,Fer07} and non-Hermitian regimes~\cite{Ros18}.

Although a vast literature on families of solvable stationary systems is available, the time-dependent counterparts have not been widely explored. The difficulty lies in the dynamical law, the Schr\"odinger equation, which is defined in terms of a partial differential equation that, in general, can not be reduced to an ordinary differential equation. Under some circumstances, we can extract information of the system through approximation techniques such as the sudden and the adiabatic approximations~\cite{Sch02}. The latter restricts the range of applicability of the so-obtained solutions. Despite all these difficulties, time-dependent phenomena find exciting applications in physical systems such as electromagnetic traps of charged particles~\cite{Pau90,Com86,Pri83,Gla92}, plasma physics~\cite{Mil19}, and in optical-analogs under the paraxial approximation~\cite{Cru17,Raz19,Con19}. 

In contradistinction to the stationary cases, the lack of an eigenvalue equation in time-dependent systems prevents us from implementing the Darboux-transformation directly, and some workarounds are in order. The latter has been addressed by Bargov-Samsonov~\cite{Bag95,Bag96}, where some intertwining operators allow us to relate an exactly solvable Schr\"odinger equation with another unknown one. In analogy to the stationary Darboux transformation, the solutions of the new models are inherited from the former one. Let us mention that orthogonality is no longer a property that can be taken for granted~\cite{Zel17}, since the method by itself does not provide essential information about the system such as the constants of motion, which have to be determined separately. Despite such a difficulty, several new families of exactly-solvable time-dependent potentials have been reported in the literature~\cite{Zel17,Con17,Cru19,Cru20}. 

Among nonstationary quantum systems, the parametric oscillator~\cite{Lew69,Dod95,Zel19,Ram18} is perhaps the most well-known model that admits a set of exact solutions. Lewis and Riesenfeld~\cite{Lew69} addressed the problem by noticing the existence of a nonstationary eigenvalue equation associated with the appropriate \textit{constant of motion} (\textit{quantum invariant}) of the system in which the time dependence appears in the coefficients of the related ordinary differential equation. The latter eigenvalue equation can indeed be factorized in such a way that the Darboux transformation\footnote{An eigenvalue equation for the time-dependent Hamiltonian is still attainable in the context of the \textit{adiabatic approximation}~\cite{Sch02}.} is applied with ease~\cite{Zel19b,Zel20}, resulting in a new quantum invariant rather than a Hamiltonian. Then, the appropriate ansatz allows to determine the respective Hamiltonian and time-dependent potentials with ease~\cite{Zel19b}. The solutions, and the complex-phases introduced by Lewis-Riesenfeld, are inherited from the former system, ensuring an orthogonal set of solutions for the new system. 

In this work, we combine the solutions of the Ermakov and fourth Painlev\'e equation to address the construction of new time-dependent Hamiltonians. This is achieved by considering third-order intertwining operators in the spatial variable with time-dependent coefficients. Those operators generate a third-order shape-invariant condition with respect to an unknown quantum invariant, which is introduced as a deformation of the one associated with the parametric oscillator. In this form, by working with a quantum invariant rather than a Hamiltonian, we generalize the construction presented in~\cite{And00,Mar09}, and the time dependence is introduced into the intertwining operators through the solutions of the Ermakov equation, which ensure the regularity of the resulting quantum invariant and its eigenfunctions at each time. On the other hand, with aid of the appropriate reparametrization, the fourth Painlev\'e equation is achieved, the parameters and solutions of which determine the spectral information and the exact form of the quantum invariant. Then, we modify the \textit{transitionless tracking algorithm}~\cite{Ber09} to construct the time-dependent Hamiltonians from the quantum invariant, from where the respective solutions of the Schr\"odinger equation are determined with the addition of a time-dependent complex-phase.

The text is structured as follows. In Sec.~\ref{sec:Painleve}, we introduce the basic notions of shape-invariance for time-dependent Hamiltonians and their respective quantum invariants. Then, a couple of differential ladder operators of third-order are introduced such that an initial, and unknown, quantum invariant satisfies a higher-order shape-invariant relationship. From the latter, the explicit form of the ladder operators and the quantum invariants are determined. In Sec.~\ref{sec:specI1I2}, with the aid of the third-order ladder operators, we determine the respective spectral information of the quantum invariant. In Sec.~\ref{sec:timeH}, the time-dependent Hamiltonians associated with the quantum invariants are identified, together with the solutions to the Schr\"odinger equation. In Sec.~\ref{sec:Ermakov}, we discuss the solutions of the Ermakov equation for some specific time-dependent frequency profiles. In particular, it is shown that the constant-frequency case leads to periodic potentials, whose solutions are in agreement with the Floquet theorem. Also, the appropriate limit to recover the well-known stationary results is presented. In turn, in Sec.~\ref{sec:solPainleve}, we consider some particular solutions of the Painlev\'e equation obtained through solutions of the Riccati equation, in the form of rational solutions, or by solutions of another nonlinear models such as the derivative nonlinear Schr\"odinger equation. For completeness, in App.~\ref{sec:PO}, we briefly revisit the parametric oscillator and its solutions through the approach of Lewis-Riesenfeld.

\section{Time-dependent quantum invariants, third-order ladder operators and the fourth Painlev\'e equation}
\label{sec:Painleve}
In quantum mechanics, the quantum invariants play a fundamental role in determining the exact solutions of the quantum models. For stationary systems (time-independent Hamiltonians), it is straightforward to realize that the Hamiltonian is a constant of motion, such that it leads to an eigenvalue equation in the form of a Sturm-Liouville problem. For time-dependent Hamiltonians, the determination of such quantum invariants becomes a challenging task in most of the cases. A prime example is given by the parametric oscillator, where the respective quantum invariants are determined with relative ease. As pointed out in~\cite{Lew69}, such an invariant admits a nonstationary eigenvalue equation, where its spectral information leads to the solutions of the Schr\"odinger equation. For the sake of self-consistency, in App.~\ref{sec:PO} we provide a brief discussion on the matter.

Thoughout this manuscript, we construct new time-dependent models and their respective solutions using a new approach based on the ladder operator structure associated with the quantum invariant. The method relies on the existence of an unknown quantum invariant $\hat{I}_{1}(t)$, which admits a set of ladder operators $\{ \hat{A}(t), \hat{A}^{\dagger}(t) \}$ defined in coordinate representation as $N$-order differential operators in the spatial variable with time-dependent coefficients. In particular, to reduce the possible family of quantum invariants, we consider $\hat{I}_{1}(t)$ as a deformation of the invariant associated with the parametric oscillator, $\hat{I}_{0}(t)$. Thus, for a fixed order $N$, we determine the exact form of $\hat{I}_{1}(t)$. It is worth to mention that ladder operators of first and second-order have been reported for the parametric oscillator~\cite{Zel19,Zel19b} and the nonstationary singular oscillator~\cite{Zel20,Dod98}. Nevertheless, in those cases, the quantum invariants are already known, and the ladder operators are constructed following the polynomial structure of the eigenfunctions. Throughout the rest of the manuscript, we focus in third-order differential ladder operators and the related quantum invariants. In this case, we determine several families of time-dependent Hamiltonians $\hat{H}_{1}(t)$, together with the respective solutions to the Schr\"odinger equation.

Before proceeding, we would like to stress the meaning of shape-invariance in supersymmetric quantum mehcnics (SUSYQM) for both stationary and nonstationary systems. It is said that two time-independent Hamiltonians $\hat{H}_{\pm}$ are shape-invariant~\cite{Kha93} if their respective potentials $V_{+}(x;\{c_{n}\})$ and $V_{-}(x;\{d_{n}\})$, with $\{c_{n}\}$ and $\{d_{n}\}$ two sets of constant parameters, are related by the condition $V_{+}(x;\{c_{n}\})=V_{-}(\{d_{n}\})+S(\{c_{n}\})$, where $S(\{c_{n}\})$ is a function of the set of parameters $\{c_{n}\}$ and independent of $x$. In turn, for nonstationary systems, we define the shape-invariance considering two quantum invariants $\hat{I}_{\pm}(t)$ by means of the relationship $\hat{I}_{+}(t)=\hat{I}_{-}(t)+f_{0}$, with $f_{0}$ a real constant. The latter implies that the respective time-dependent Hamiltomnians $\hat{H}_{\pm}(t)$ are related as $\hat{H}_{+}(t)=\hat{H}_{-}(t)+f(t)$, with $f(t)$ an arbitrary real-valued function of time. As a consequence, the solutions of the respective Schr\"odinger equations differ only from a global time-dependent complex-phase $e^{i\int^{t} dt'f(t')}$, see~\cite{Zel19b} for details. 

From the previous considerations, let us introduce two unknown quantum invariants $\hat{I}_{1,2}(t)$, where $\hat{I}_{1}(t)$ is the invariant operator under consideration, and $\hat{I}_{2}(t)$ serves as an auxiliary operators to determined $\hat{I}_{1}(t)$. From the latter, there are two time-dependent Hamiltonians $\hat{H}_{1,2}(t)$ such that
\begin{equation}
\frac{d\hat{I}_{j}(t)}{dt}=i\left[\hat{H}_{j}(t),\hat{I}_{j}(t) \right]+\frac{\partial \hat{I}_{j}(t)}{\partial t}=0 \, , \quad j=1,2 \, .
\end{equation}
We focus on self-adjoint time-dependent Hamiltonians and quantum invariants. It is thus guaranteed that the nonstationary eigenvalue equations
\begin{equation}
\hat{I}_{j}(t)\phi^{(j)}_{n}(x,t)=\Lambda^{(j)}_{n}\phi^{(j)}_{n}(x,t) \, , \quad j=1,2,
\label{eq:specI1}
\end{equation}
lead to real and time-independent eigenvalues $\Lambda^{(j)}_{n}$, and orthogonal nonstationary eigenfunctions $\phi^{(j)}_{n}(x,t)$ that satisfy the finite-norm condition $\vert\langle \phi^{(j)}_{n}(t)\vert\phi^{(j)}_{n}(t)\rangle\vert<\infty$. The orthogonality condition can be used as a base to construct the vector spaces $\mathcal{H}_{j}(t)=Span\{\phi^{(j)}_{n}(x,t)\}_{n=0}^{\infty}$, with $j=1,2$. 

As mentioned above, we construct the quantum invariant $\hat{I}_{1}(t)$ from the third-order SUSYQM shape-invariant condition, defined in term of the intertwining relationships
\begin{equation}
\hat{I}_{1}(t)\hat{A}^{\dagger}(t)=\hat{A}^{\dagger}(t)\left[ \hat{I}_{1}(t)+2\lambda \right] \, , \quad \hat{I}_{1}(t)\hat{A}(t)=\hat{A}(t)\left[ \hat{I}_{1}(t)-2\lambda \right] \, ,
\label{eq:SINV1}
\end{equation}
where $\hat{A}(t)$ and $\hat{A}^{\dagger}(t)$ are also the annihilation and creation operators, respectively, for $\phi_{n}^{(1)}(x,t)$. Indeed, the action of $\hat{A}^{\dagger}(t)$ ($\hat{A}(t)$) on $\phi_{n}^{(1)}(x,t)$ increases (decreases) the eigenvalue $\Lambda_{n}^{(1)}$ by $2\lambda$ units.

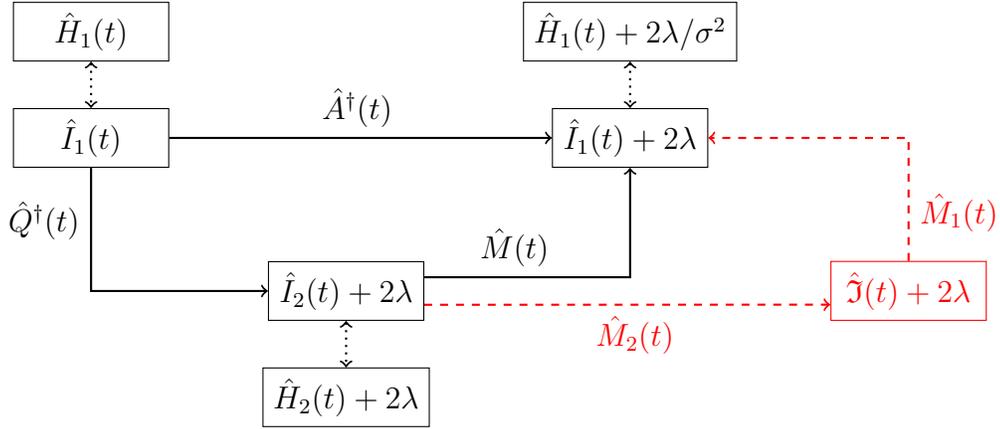
\begin{figure}
\centering
\begin{tikzpicture}
\tikzstyle{line} = [draw, thick, -latex’,shorten >=2pt];
\matrix (m) [matrix of nodes,row sep=1.5em,column sep=3em,minimum width=5em, nodes={draw}]
{
\node(m11) {$\hat{H}_{1}(t)$};	& 			& \node(m12) {$\hat{H}_{1}(t)+2\lambda/\sigma^{2}$}; & \\     
\node(m21) {$\hat{I}_{1}(t)$};	& 	 	    & \node(m22) {$\hat{I}_{1}(t)+2\lambda$}; & \\
     	    &    		&   				  & \\
 & \node(m32) {$\hat{I}_{2}(t)+2\lambda$}; & & \node(m34)[red] {$\hat{\mathfrak{I}}(t)+2\lambda$}; \\
 & \node(m42) {$\hat{H}_{2}(t)+2\lambda$}; & & \\};
\draw [<->,thick,dotted] (m11.south) -- (m21.north);
\draw [<->,thick,dotted] (m12.south) -- (m22.north);
\draw [<->,thick,dotted] (m32.south) -- (m42.north);
\draw [->,black,thick] (m21.east) node[above,xshift=2.5cm] {$\hat{A}^{\dagger}(t)$} -- (m22.west);
\draw [->,black,thick] (m21.south) node[left,yshift=-0.7cm] {$\hat{Q}^{\dagger}(t)$} |- (m32.west);
\draw [->,black,thick] (m32.10) node[above,xshift=1.2cm] {$\hat{M}(t)$} -| (m22.south);
\draw [->,red,dashed,thick] (m32.-10) node[below,xshift=2.8cm] {$\hat{M}_{2}(t)$} -- (m34.-170);
\draw [->,red,dashed,thick] (m34.north) node[right,yshift=0.65cm] {$\hat{M}_{1}(t)$} |- (m22.east);
\end{tikzpicture}
\caption{\footnotesize{Third-order shape-invariant SUSYQM for the quantum invariant $\hat{I}_{1}(t)$. The arrow indicates intertwining relationship between the quantum invariants; for instance, the arrow on top implies $\hat{I}_{1}(t)\hat{A}^{\dagger}(t)=\hat{A}^{\dagger}[\hat{I}_{2}(t)+2\lambda]$. The direction of arrows is inverted by using the adjoint relationships. Red lines show the reducible factorization discussed in Sec.~\ref{sec:specI1I2}. The Hamiltonians $\hat{H}_{1,2}(t)$ are presented in Sec.~\ref{sec:timeH}.}}
\label{fig:F0}
\end{figure}

Now, to provide a specific form to the quantum invariants $\hat{I}_{1,2}(t)$, we consider them as deformations of the quantum invariant associated with the parametric oscillator $\hat{I}_{0}(t)$, see Eq.~\eqref{eq:invI0}, that is,
\begin{equation}
\hat{I}_{j}(t)=\hat{I}_{0}(t)+R_{j}(\hat{x},t)\equiv -\sigma^{2}\frac{\partial^{2}}{\partial x^{2}}+ix\sigma\dot{\sigma}\frac{\partial}{\partial x} + R(x,t) + R_{j}(x,t) \, , \quad j=1,2,
\label{eq:INVI1}
\end{equation}
where $\sigma\equiv\sigma(t)$ given in~\eqref{eq:nonlinear} is solution to the Ermakov equation associated with the parametric oscillator~\eqref{eq:ermakov}, and $\dot{\sigma}\equiv d\sigma/dt$. The function $R(x,t)$ is given as
\begin{equation}
R(x,t)=\left(\frac{\dot{\sigma}^{2}}{4}+\frac{1}{\sigma^{2}}\right)x^{2}+i\frac{\dot{\sigma}\sigma}{2} \, ,
\label{eq:R}
\end{equation}
and $R_{1,2}(x,t)$ are real-valued functions, where $R_{1}(x,t)$ is determined from the shape-invariant condition~\eqref{eq:SINV1}, and $R_{2}(x,t)$ defines the auxiliary invariant $\hat{I}_{2}(t)$. As discussed in~\cite{And00}, higher-order intertwining operators can be factorized as products of first-order operators for a reducible factorization, or as combination of first and second-order operators for irreducible factorizations~\cite{And95}. In the sequel, we focus on the reducible factorization of the intertwining relationships~\eqref{eq:SINV1}; however, for the sake of completeness, we discuss the more general irreducible factorization in this section. We thus decompose the set of intertwining operators $\{ \hat{A}(t), \hat{A}^{\dagger}(t)\}$ as the product of first and second-order differential operators $\{\hat{Q}^{\dagger}(t),\hat{Q}(t)\}$ and $\{\hat{M}^{\dagger}(t),\hat{M}(t)\}$, respectively, as follows:
\begin{equation}
\hat{A}^{\dagger}(t)=\hat{Q}^{\dagger}(t)\hat{M}(t) \, , \quad \hat{A}=\hat{M}^{\dagger}(t)\hat{Q}(t) \, .
\label{eq:factA}
\end{equation}
The new operators give rise the additional set of intertwining relationships of the form
\begin{equation}
\hat{I}_{1}(t)\hat{Q}^{\dagger}(t)=\hat{Q}^{\dagger}(t)\left[ \hat{I}_{2}(t)+2\lambda \right] \, , \quad \hat{I}_{2}(t)\hat{Q}(t)=\hat{Q}(t)\left[ \hat{I}_{1}(t)-2\lambda \right] \, ,
\label{eq:INTER1}
\end{equation}
\begin{equation}
\hat{I}_{2}(t)\hat{M}(t)=\hat{M}(t)\hat{I}_{1}(t) \, , \quad \hat{I}_{1}(t)\hat{M}^{\dagger}(t)=\hat{M}^{\dagger}(t)\hat{I}_{2}(t) \, ,
\label{eq:INTER2}
\end{equation}
where in the latter we have introduced the auxiliary quantum invariant $\hat{I}_{2}(t)$ as an intermediate. In contradistinction to~\eqref{eq:SINV1}, the Eqs.~\eqref{eq:INTER1}-\eqref{eq:INTER2} by themselves do not define a shape-invariant relation. Nevertheless, their combined action take us back to the shape-invariant condition~\eqref{eq:SINV1}, see Fig.~\ref{fig:F0} for details. 

Now, given that $\hat{Q}(t)$ and $\hat{Q}^{\dagger}(t)$ are considered as first-order differential operators, we use the general form introduced in~\cite{Zel20}, that is,
\begin{equation}
\hat{Q}^{\dagger}=\sigma\frac{\partial}{\partial x}+\mathfrak{w}(x,t) \, , \quad \hat{Q}=-\sigma\frac{\partial}{\partial x}+\mathfrak{w}^{*}(x,t) \, ,
\label{eq:FACQ}
\end{equation}
with $\mathfrak{w}(x,t)$ a complex-valued function, $\sigma\equiv\sigma(t)$ given in~\eqref{eq:nonlinear}, and $f^{*}$ stands for the complex-conjugate of $f$. In turn, the second-order differential operators $\hat{M}(t)$ and $\hat{M}^{\dagger}(t)$ are constructed as a generalization of those reported in~\cite{And95} by introducing the time-dependent coefficients of the form 
\begin{equation}
\begin{aligned}
& \hat{M}^{\dagger}=\sigma^{2}\frac{\partial^{2}}{\partial x^{2}}-2 g(x,t)\frac{\partial}{\partial x}+b(x,t) \, , \\
& \hat{M}=\sigma^{2}\frac{\partial^{2}}{\partial x^{2}}+2g^{*}(x,t)\frac{\partial}{\partial x}+b^{*}(x,t)-2\left[g'(x,t)\right]^{*} \, ,
\end{aligned}
\label{eq:FACM}
\end{equation}
where $b(x,t)$ and $g(x,t)$ are complex-valued functions, and $g'(x,t)$ stands for the partial derivative of $g(x,t)$ with respect to $x$. 

The complex-valued functions $\mathfrak{w}(x,t)$, $g(x,t)$ and $b(x,t)$ are determined from the respective intertwining relationships. For instance, $\mathfrak{w}(x,t)$ is obtained after substituting~\eqref{eq:FACQ} in~\eqref{eq:INTER1}, leading to
\begin{equation}
\mathfrak{w}(x,t)=-i\frac{\dot{\sigma}}{2}x+W(z(x,t)) \, , \quad z(x,t):=\frac{x}{\sigma} \, .
\label{eq:w1}
\end{equation}
Given that the solution to the Ermakov equation $\sigma(t)$ is a nodeless function, we can guarantee that the reparametrizated variable $z(x,t)$ is non-singular for $t\in\mathbb{R}$. In turn, $W(z(x,t))$ is a real-valued function that solves the Riccati equations
\begin{equation}
z^{2}+R_{1}(z)=\partial_{z}W+W^{2} \, , \quad z^{2}+R_{2}(z)=-\partial_{z}W+W^{2}-2\lambda \, , \quad \partial_{z}\equiv\frac{\partial}{\partial z} \, .
\label{eq:w2}
\end{equation}
From~\eqref{eq:w2} we also get $R_{2}(z)-R_{1}(z)=-2\partial_{z}W-2\lambda$, which resembles the conventional relationship between the potential and the super-potential of the conventional stationary SUSYQM construction~\cite{And00,Mar09}. 

On the other hand, the functions $g(x,t)$ and $b(x,t)$ are computed after inserting~\eqref{eq:FACM} in~\eqref{eq:INTER2}, leading to 
\begin{equation}
\begin{aligned}
& g(x,t)=i\frac{\sigma\dot{\sigma}}{2}x+\sigma G(z(x,t)) \, , \\
& b(x,t)= i \dot{\sigma} x G(z(x,t))-i\frac{\dot{\sigma}\sigma}{2}-\frac{\dot{\sigma}^{2}}{4}x^{2}+B(z(x,t)) \, ,
\end{aligned}
\label{eq:gb}
\end{equation}
with the real-valued functions $G(z(x,t))$ and $B(z(x,t))$ determined from the nonlinear relationships
\begin{equation}
\begin{aligned}
& B=2G^{2}+\partial_{z}G-\left( z^{2} + R_{2}(z) \right)+\gamma \, ,   \\
& z^{2}+R_{1}(z)= -2 \partial_{z}G+G^{2}+\frac{\partial^{2}_{z}G}{2G}-\frac{(\partial_{z}G)^{2}}{4G^{2}}-\frac{d}{4G^{2}}+\gamma \, , \\
& z^{2}+R_{2}(z)= 2 \partial_{z}G+G^{2}+\frac{\partial^{2}_{z}G}{2G}-\frac{(\partial_{z}G)^{2}}{4G^{2}}-\frac{d}{4G^{2}}+\gamma \, ,
\end{aligned}
\label{eq:GB}
\end{equation}
with $\gamma$ and $d$ constants of integration with respect to $z(x,t)$, that is, those constants do not depend on $x$ or $t$. From~\eqref{eq:GB} we obtain a complementary relationship of the form $R_{2}(z)-R_{1}(z)=4\partial_{z}G$ that, together with the one obtained from~\eqref{eq:w2}, gives
\begin{equation}
W(z)=-2G(z)-\lambda z \, .
\label{eq:ANSW}
\end{equation}
A differential equation for $G(z)$ can be found after substituting~\eqref{eq:ANSW} into any of the Riccati equations in~\eqref{eq:w2} and comparing with ~\eqref{eq:GB}. The straightforward calculation leads to
\begin{equation}
\partial^{2}_{z}G=\frac{(\partial_{z}G)^{2}}{2G}+6G^{3}+8\lambda z G^{2}+2\left[ \lambda^{2}z^{2}-(\gamma+\lambda) \right]G+\frac{d}{2G} \, ,
\label{eq:PAIN1}
\end{equation}
where the following reparametrizations:
\begin{equation}
y=\sqrt{\lambda}z \, , \quad G=\frac{\sqrt{\lambda}}{2}w(y) \, ,\quad \alpha=\frac{\gamma}{\lambda}+1 \, , \quad \beta=\frac{2d}{\lambda^{2}} \, ,
\label{eq:GREP}
\end{equation}
allow us to rewrite~\eqref{eq:PAIN1} as the fourth Painlev\'e differential equation~\cite{Olv10}
\begin{equation}
w_{yy}=\frac{(w_{y})^{2}}{2w}+\frac{3}{2}w^{3}+4yw^{2}+2(y^{2}-\alpha)w+\frac{\beta}{w} \, .
\label{eq:painleveIV}
\end{equation}
Solutions for the fourth Painlev\'e equation have been extensively studied in the literature, in particular it is known that $w(y)$ can be determined in terms of elementary functions~\cite{Gro02,Mar06}.

Before finishing this section, we would like to recall that the factorization~\eqref{eq:factA} allowed us to find the functions $R_{1,2}(x,t)$, which define uniquely the respective quantum invariants $\hat{I}_{1,2}(t)$ in terms of the solutions of the fourth Painlev\'e equation~\eqref{eq:painleveIV}. A summary of the steps followed so far is presented in the diagram of Fig.~\ref{fig:F0}, where the time-dependent Hamiltonians $\hat{H}_{1,2}(t)$ are discussed in Sec.~\ref{sec:timeH}.

\section{Spectral information of $\hat{I}_{1}(t)$}
\label{sec:specI1I2}
As pointed out in the previous section, the shape-invariant condition~\eqref{eq:SINV1} implies that $\hat{A}(t)$ and $\hat{A}^{\dagger}(t)$ are the ladder operators for the nonstationary eigenfunctions $\phi_{n}^{(1)}(x,t)$ of $\hat{I}_{1}(t)$. The latter indeed allows determining the spectral information~\eqref{eq:specI1}, for $j=1$. To this end, we first determine the \textit{zero-mode} eigenfunction, which is an element in the kernel of the annihilation operator $\mathcal{K}_{A}\equiv Ker(\hat{A}(t))=\{\phi^{(1)}\}$, with $\hat{A}(t)\phi^{(1)}=0$. However, in our case, the annihilation operator under consideration is a differential third-order one, and thus $\mathcal{K}_{A}$ is composed of three linearly independent zero-mode solutions, $\mathcal{K}_{A}=\{\phi^{(1)}_{0;1},\phi^{(1)}_{0;2},\phi^{(1)}_{0;3}\}$. Nevertheless, we must verify whether the elements in $\mathcal{K}_{A}$ fulfill the finite-norm condition. With the zero-modes already identified, the remaining eigenfunctions are computed from the iterated action of the creation operator $\hat{A}^{\dagger}(t)$ on the zero-mode eigenfunctions, and the respective eigenvalues increase by $2\lambda$ at each iteration.

For convenience, in this section we consider the case for which $\hat{M}(t)^{\dagger}$ factorizes as the product of two first-order operators, that is, a \textit{reducible} case. Let us consider the factorization
\begin{equation}
\hat{M}^{\dagger}(t)\equiv\hat{M}_{1}^{\dagger}(t)\hat{M}_{2}^{\dagger}(t) \, , \quad \hat{M}(t)=\hat{M}_{2}(t)\hat{M}_{1}(t) \, ,
\end{equation}
where $\hat{M}_{1,2}(t)$ are first-order operators constructed in analogy to~\eqref{eq:FACQ} as
\begin{equation}
\hat{M}_{1}^{\dagger}(t):=\left( \sigma\frac{\partial}{\partial x}-i\frac{\dot{\sigma}}{2}x+W_{1}(z) \right) \, , \quad \hat{M}_{2}^{\dagger}(t):=\left( \sigma\frac{\partial}{\partial x}-i\frac{\dot{\sigma}}{2}x+W_{2}(z) \right) \, .
\label{eq:M1M2}
\end{equation}
The straightforward calculations show that the real-valued functions $W_{1}(z)$ and $W_{2}(z)$ are given by
\begin{equation}
W_{1}=-G+\left(\frac{G_{z}-\sqrt{-d}}{2G}\right) \, , \quad W_{2}=-G-\left(\frac{G_{z}-\sqrt{-d}}{2G}\right) \, .
\label{eq:W1W2}
\end{equation}
From the latter result, it is clear that the factorization of $\hat{M}(t)$ requires $d<0$. Recall that $\hat{M}^{\dagger}(t)$ intertwines the quantum invariant $\hat{I}_{1}(t)$ with $\hat{I}_{2}(t)$. Thus, to inspect the respective intertwining relationships fulfilled by $\hat{M}_{1,2}(t)$ we introduce a new auxiliary quantum invariant
\begin{equation}
\hat{\frak{I}}\tilde{\phi}_{n}(x,t)=\tilde{\Lambda}_{n}\tilde{\phi}_{n}(x,t)
\label{eq:Itilde}
\end{equation}
with $\tilde{\mathcal{H}}=Span\{ \tilde{\phi}_{n} \}_{n=0}^{\infty}$ the respective vector space composed with the finite-norm solutions. The spectral information of $\hat{\mathfrak{I}}(t)$ is not relevant, for it just serves as an aid to solve the eigenvalue problem associated with $\hat{I}_{1}(t)$. For this reason, the respective Hamiltonian associated with $\hat{\mathfrak{I}}(t)$ is not considered throughout the rest of the text.

The new auxiliary invariant satisfies the intertwining relationships
\begin{equation}
\hat{I}_{1}(t)\hat{M}_{1}^{\dagger}(t)=\hat{M}_{1}^{\dagger}(t)\hat{\mathfrak{I}}(t) \, , \quad \hat{\mathfrak{I}}(t)\hat{M}_{2}^{\dagger}(t)=\hat{M}_{2}^{\dagger}(t)\hat{I}_{2}(t) \, .
\label{eq:intert2}
\end{equation}
Given that both operators $\hat{M}_{1,2}(t)$ are of first-order, the relationships~\eqref{eq:intert2} are then equivalent to
\begin{equation}
\begin{alignedat}{3}
& \hat{I}_{1}(t)=\hat{M}_{1}^{\dagger}(t)\hat{M}_{1}(t)+\epsilon_{1} \, , \quad && \hat{\mathfrak{I}}(t)=\hat{M}_{1}(t)\hat{M}_{1}^{\dagger}(t)+\epsilon_{1} \, , \\
& \hat{\mathfrak{I}}(t)=\hat{M}^{\dagger}_{2}(t)\hat{M}_{2}(t)+\epsilon_{2} \, , && \hat{I}_{2}(t)=\hat{M}_{2}(t)\hat{M}_{2}^{\dagger}(t)+\epsilon_{2} \, ,
\end{alignedat}
\label{eq:facI1I2}
\end{equation}
where the substitution of~\eqref{eq:M1M2}-\eqref{eq:W1W2} into~\eqref{eq:facI1I2} leads to
\begin{equation}
\epsilon_{1}=\gamma-\sqrt{-d} \, , \quad \epsilon_{2}=\gamma+\sqrt{-d} \, .
\label{eq:e1e2}
\end{equation}
From~\eqref{eq:INTER1} and~\eqref{eq:intert2}, we can see that the first-order operators define mappings among the vector spaces $\mathcal{H}_{1}(t)$, $\mathcal{H}_{2}(t)$ and $\tilde{\mathcal{H}}(t)$ in the following form:
\begin{equation}
\begin{alignedat}{3}
& \hat{M}_{2}^{\dagger}(t):\mathcal{H}_{2}(t)\rightarrow\tilde{\mathcal{H}}(t) \, , \quad &&\hat{M}_{2}(t):\tilde{\mathcal{H}}(t)\rightarrow\mathcal{H}_{2}(t) \, , \\
& \hat{M}_{1}^{\dagger}(t):\tilde{\mathcal{H}}(t)\rightarrow\mathcal{H}_{1}(t)\, , \quad && M_{1}(t):\mathcal{H}_{1}(t)\rightarrow\tilde{\mathcal{H}}(t) \, , \\
& \hat{Q}^{\dagger}(t):\mathcal{H}_{2}(t)\rightarrow\mathcal{H}_{1}(t)\, , \quad && Q(t):\mathcal{H}_{1}(t)\rightarrow\mathcal{H}_{2}(t) \, .
\end{alignedat}
\end{equation}

From the latter mappings, we construct the elements of $\mathcal{K}_{A}$ (zero-modes), that is, the eigenfunctions annihilated by $\hat{A}(t)$. We thus have
\begin{equation}
\hat{A}(t)\phi_{0;k}^{(1)}=\hat{M}^{\dagger}(t)\hat{Q}(t)\phi_{0;k}^{(1)}=\hat{M}_{1}^{\dagger}\hat{M}_{2}^{\dagger}(t)\hat{Q}(t)\phi_{0;k}^{(1)}=0 \, , \quad k=1,2,3,
\label{eq:kernelA}
\end{equation}
where it is worth discussing three different cases.

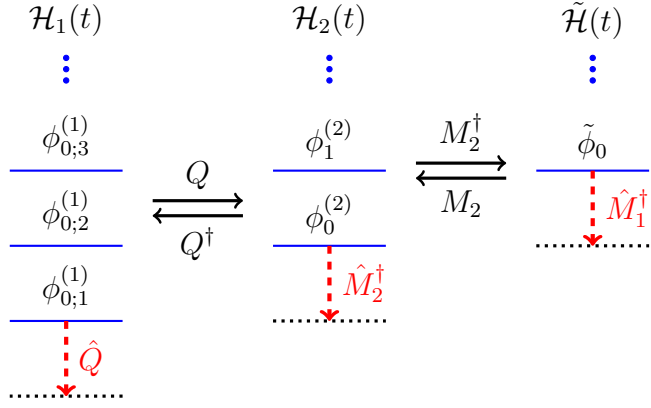
\begin{figure}
\centering
\begin{tikzpicture}
\draw (0.75,3) node {$\mathcal{H}_{1}(t)$};
\draw (4.25,3) node {$\mathcal{H}_{2}(t)$};
\draw (7.75,3) node {$\tilde{\mathcal{H}}(t)$};
\filldraw[blue] (0.75,2.2) circle (1pt);
\filldraw[blue] (0.75,2.35) circle (1pt);
\filldraw[blue] (0.75,2.5) circle (1pt);
\filldraw[blue] (4.25,2.2) circle (1pt);
\filldraw[blue] (4.25,2.35) circle (1pt);
\filldraw[blue] (4.25,2.5) circle (1pt);
\filldraw[blue] (7.75,2.2) circle (1pt);
\filldraw[blue] (7.75,2.35) circle (1pt);
\filldraw[blue] (7.75,2.5) circle (1pt);
\draw[blue,thick] (1.5,1)--(0,1) node[midway,anchor=south,text=black] {$\phi_{0;3}^{(1)}$};
\draw[blue,thick] (1.5,0)--(0,0) node[midway,anchor=south,text=black] {$\phi_{0;2}^{(1)}$};
\draw[blue,thick] (1.5,-1)--(0,-1) node[midway,anchor=south,text=black] {$\phi_{0;1}^{(1)}$};
\draw[black,very thick,dotted] (1.5,-2)--(0,-2);
\draw[red,ultra thick,dashed,->] (0.75,-1)--(0.75,-2) node[midway,anchor=west,text=red] {$\hat{Q}$};
\draw[blue,thick] (3.5,1)--(5,1) node[midway,anchor=south,text=black] {$\phi_{1}^{(2)}$};
\draw[blue,thick] (3.5,0)--(5,0) node[midway,anchor=south,text=black] {$\phi_{0}^{(2)}$};
\draw[black,very thick,dotted] (3.5,-1)--(5,-1);
\draw[red,ultra thick,dashed,->] (4.25,0)--(4.25,-1) node[midway,anchor=west,text=red] {$\hat{M}_{2}^{\dagger}$};
\draw[blue,thick] (7,1)--(8.5,1) node[midway,anchor=south,text=black] {$\tilde{\phi}_{0}$};
\draw[black,very thick,dotted] (7,0)--(8.5,0);
\draw[red,ultra thick,dashed,->] (7.75,1)--(7.75,0) node[midway,anchor=west,text=red] {$\hat{M}_{1}^{\dagger}$};
\draw[very thick,->] (1.9,0.6)--(3.1,0.6) node[midway,anchor=south] {$Q$};
\draw[very thick,->] (3.1,0.4)--(1.9,0.4) node[midway,anchor=north] {$Q^{\dagger}$};
\draw[very thick,->] (5.4,1.1)--(6.6,1.1) node[midway,anchor=south] {$M_{2}^{\dagger}$};
\draw[very thick,->] (6.6,0.9)--(5.4,0.9) node[midway,anchor=north] {$M_{2}$};
\end{tikzpicture}
\caption{\footnotesize{Zero-mode eigenfunctions $\{ \phi_{0;1}^{(1)}, \phi_{0;2}^{(1)}, \phi_{0;3}^{(1)} \}$ and their relationship with the respective eigenfunctions of the auxiliary quantum invariants $\hat{I}_{2}(t)$ and $\hat{\mathfrak{I}}(t)$. Dotted-black lines denote the null-vector in vector space, and the dashed-red arrows indicate the annihilation operation of the intertwining operators.}}
\label{fig:F2}
\end{figure}

$\bullet$ $\hat{Q}(t)\phi_{0;1}^{(1)}=0$. Here, a first solution is determined, up to a normalization constant, by solving a trivial first-order differential equation.

$\bullet$ $\hat{M}_{2}^{\dagger}(t)\hat{Q}(t)\phi_{0;2}^{(1)}= 0$ with $\hat{Q}(t)\phi_{0;2}^{(1)}\neq 0$. From the mappings defined by the intertwining relationship~\eqref{eq:INTER1}, it is clear that $\hat{Q}(t)\phi_{0;2}^{(1)}=\phi_{0}^{(2)}\in\mathcal{H}_{2}(t)$, with the latter being annihilated by $\hat{M}_{2}^{\dagger}(t)$. Thus, in order to determine the zero-mode $\phi_{0;2}^{(1)}$, we should solve the first-order differential equation $\hat{M}_{2}^{\dagger}(t)\phi_{0}^{(2)}=0$, and from it we determine $\phi_{0;2}^{(1)}$, together with the respective eigenvalue, after mapping $\phi_{0}^{(2)}$ through $\hat{Q}^{\dagger}(t)$ (see Fig.~\ref{fig:F2}).

$\bullet$ $\hat{M}_{1}^{\dagger}(t)\hat{M}_{2}^{\dagger}(t)\hat{Q}(t)\phi_{0;3}^{(1)}= 0$ with $\hat{M}_{2}^{\dagger}(t)\hat{Q}(t)\phi_{0;3}^{(1)}\neq 0$. In this case, the non-null element $\hat{M}_{2}^{\dagger}(t)\hat{Q}(t)\phi_{0;3}^{(1)}=\tilde{\phi}_{0}\in\tilde{\mathcal{H}}(t)$ is annihilated by $\hat{M}_{1}^{\dagger}(t)$. To extract the zero-mode $\phi_{0;3}^{(1)}$, we solve the first-order differential equation $\hat{M}_{1}^{\dagger}(t)\tilde{\phi}_{0}=0$. Then, we take $\tilde{\phi}_{0}$ to $\mathcal{H}_{1}(t)$ by consecutively performing the mappings $\hat{M}_{2}(t)$ and $\hat{Q}^{\dagger}(t)$ (see Fig.~\ref{fig:F2}). 

The complete procedure is summarized in the scheme depicted in Fig.~\ref{fig:F2}. The straightforward calculations lead to
\begin{equation}
\begin{aligned}
& \phi_{0;1}^{(1)}(x,t):=\mathcal{N}_{0;1}^{(1)} \, \frac{e^{\frac{i}{4}\frac{\dot{\sigma}}{\sigma}x^{2}}}{\sqrt{\sigma}}e^{\int^{z} dz' W(z')} \, , \\
& \phi_{0;2}^{(1)}(x,t)=\mathcal{N}_{0;1}^{(1)} \left[ W(z)-W_{2}(z) \right] \frac{e^{\frac{i}{4}\frac{\dot{\sigma}}{\sigma}x^{2}}}{\sqrt{\sigma}}e^{-\int^{z} dz' W_{2}(z')} \, , \\
& \phi_{0;3}^{(1)}(x,t)=\mathcal{N}_{0;3}^{(1)}\left[ -2\sqrt{-d}+(W(z)-W_{2}(z))(W_{1}(z)+W_{2}(z))\right]\frac{e^{\frac{i}{4}\frac{\dot{\sigma}}{\sigma}x^{2}}}{\sqrt{\sigma}}e^{-\int^{z} dz' W_{1}(z')} \, , 
\end{aligned}
\label{eq:ground}
\end{equation}
with the respective eigenvalues $\Lambda_{0;1}^{(1)}=0$, $\Lambda_{0;2}^{(1)}=\epsilon_{2}+2\lambda=\gamma+\sqrt{-d}+2\lambda$ and $\Lambda_{0;3}^{(1)}=\epsilon_{1}+2\lambda=\gamma-\sqrt{-d}+2\lambda$. The terms $\mathcal{N}_{0;j}^{(1)}$ stand for the normalization factors that might depend on time.  From~\eqref{eq:ground}, the rest of the eigenfunctions are determined from the action $\left[\hat{A}^{\dagger}(t)\right]^{n}$, for $n=0,1,\cdots$, on each element $\phi_{0;j}^{(1)}$. By doing so, we generate at most three sequences of eigenfunctions, where the eigenvalues $\Lambda_{0;j}^{(1)}$, for $j=1,2,3$, increase by $2n\lambda$. 

For the conventional stationary oscillator, it is well-known that the creation operator does not lead to finite-norm eigenfunctions. On the other hand, the one-step SUSY partner Hamiltonians admit a creation operator for which a finite-norm eigenfunction is achived. In the context of SUSYQM, such an eigenfunction is the so-called \textit{missing state}~\cite{Fer99}. Thus, it is natural to look for the solutions that are annihilated by the creation operator. In the case under consideration, we have constructed $\hat{A}^{\dagger}(t)$ as a third-order differential operator, which admits three linearly independent eigenfunctions, and at least one finite-norm solution is possible. The existence of the latter implies a truncation of the sequences generated from the zero-modes $\phi_{0;j}^{(1)}$. We thus define $\mathcal{K}_{A^{\dagger}}:=Ker(\hat{A}^{\dagger}(t))=\{ \Phi_{0;1}^{(1)}, \Phi_{0;2}^{(1)}, \Phi_{0;3}^{(1)} \}$ as the set containing the finite-norm eigenfunctions of $\hat{A}^{\dagger}(t)$. If the set $\mathcal{K}_{A^{\dagger}}$ is empty, three infinite sequences are generated (see Sec.~\ref{subsec:rational}). In turn, if $\mathcal{K}_{A^{\dagger}}$ contains one single element, we generate at most two infinite sequences, together with one finite-dimensional sequence (see Sec.~\ref{subsec:nonlinb}), which in particular could be a singlet (see Sec.~\ref{subsec:riccati}).

Following the same steps as in~\eqref{eq:ground}, it is straightforward to show that the eigenfunctions of $A^{\dagger}$ are
\begin{equation}
\begin{aligned}
& \Phi_{0;1}^{(1)}(x,t)=\overline{\mathcal{N}}_{0;1}^{(1)} \, \frac{e^{\frac{i}{4}\frac{\dot{\sigma}}{\sigma}x^{2}}}{\sqrt{\sigma}}e^{\int^{z} dz' W_{1}(z')} \, , \\
& \Phi_{0;2}^{(1)}(x,t):=\overline{\mathcal{N}}_{0;2}^{(1)} \left[ W_{1}(z)+W_{2}(z)\right] \frac{e^{\frac{i}{4}\frac{\dot{\sigma}}{\sigma}x^{2}}}{\sqrt{\sigma}}e^{\int^{z} dz' W_{2}(z')} \, \\
& \Phi_{0;3}^{(1)}(x,t):= \overline{\mathcal{N}}_{0;3}^{(1)}\left[ \epsilon_{2}+2\lambda+(W_{1}(z)+W_{2}(z))(W_{2}(z)-W(z))\right]\frac{e^{\frac{i}{4}\frac{\dot{\sigma}}{\sigma}x^{2}}}{\sqrt{\sigma}}e^{-\int^{z} dz' W(z')} \, ,
\end{aligned}
\label{eq:kernelAD}
\end{equation}
where the respective eigenvalues are given by $\overline{\Lambda}_{0,1}^{(1)}=\epsilon_{1}=\gamma-\sqrt{-d}$, $\overline{\Lambda}_{0,2}^{(1)}=\epsilon_{2}=\gamma+\sqrt{-d}$ and $\overline{\Lambda}_{0,3}^{(1)}=-2\lambda$. 

In general, we can not say which solutions in~\eqref{eq:ground} and~\eqref{eq:kernelAD} fulfill the finite-norm condition, since it depends on the specific solutions of the fourth Painlev\'e equation. However, we may get more insight by considering the possible behavior of the asymptotics. To this end, let us suppose that the real-valued functions $W_{1}(z)$, $W_{2}(z)$ and $W(z)$ are smooth, and such that they converge to a finite value for $\vert z\vert\rightarrow\infty$. Then, finite-norm solutions are achieved depending on the convergence of the exponential functions in~\eqref{eq:ground}-\eqref{eq:kernelAD}. For instance, if $e^{\int^{z} dz' W(z')}\rightarrow 0$ for $\vert z\vert\rightarrow\infty$, then $\phi_{0;1}^{(1)}(x,t)$ becomes a finite-norm solution, whereas $\Phi_{0;3}^{(1)}(x,t)$ do not. The same analysis can be extended to the rest of solutions in~\eqref{eq:ground}-\eqref{eq:kernelAD}. In such a case, we may conclude that at most three out of the six solutions have a finite-norm. This indeed corresponds to the solutions discussed in Sec.~\ref{subsec:riccati} and Sec.~\ref{subsec:rational}. 


\section{New families of time-dependent Hamiltonians}
\label{sec:timeH}
So far, we have determined the families of exactly solvable quantum invariant $\hat{I}_{1}(t)$, related to the fourth Painle\'e transcendents, which fulfill a third-order SUSYQM shape-invariant condition. Nevertheless, the respective Hamiltonian of the system $\hat{H}_{1}(t)$ has not been identified yet. The latter is required to properly define the Schr\"odinger equation that characterize the quantum system under consideration. Such a task have been addressed in previous works using the factorization method for time-dependent Hamiltonians~\cite{Zel19b,Zel20} by imposing the appropriate ansatz. In this work, we consider an alternative approach based on the transitionless tracking algorithm~\cite{Ber09,Che11}. In this form, additional information is obtained about the classes of time-dependent Hamiltonians that can be constructed in term of the nonstationary eigenfunctions.

In App.~\ref{sec:PO} we have discussed the nonstationary eigenvalue equation of the quantum invariant associated with the parametric oscillator. Remarkably, the results from Lews-Riesenfeld~\cite{Lew69} hold for any quantum invariant\footnote{Although, an orthogonal set of eigenfunctions can not be taken for granted for any general quantum invariant.}. Therefore, for the quantum invariants $\hat{I}_{1,2}(t)$ constructed in Sec.~\ref{sec:Painleve} we can determine the respective time-dependent Hamiltonians $\hat{H}_{1,2}(t)$ such that the Schr\"odinger equation, in coordinate-free representation,
\begin{equation}
i\partial_{t}\vert\psi_{n}^{(j)}(t)\rangle=\hat{H}_{j}(t)\vert\psi_{n}^{(j)}(t)\rangle \, , \quad \vert\psi_{n}^{(j)}(t)\rangle=e^{i\theta_{n}^{(j)}(t)}\vert\phi_{n}^{(j)}(t)\rangle \, , \quad j=1,2,
\label{eq:schr}
\end{equation}
is fulfilled, where the wavefunctions and nonstationary eigenfunctions are recovered from the coordinate-representation $\psi_{n}^{(j)}(x,t)=\langle x\vert\psi_{n}^{(j)}(t)\rangle$ and $\phi_{n}^{(j)}(x,t)=\langle x \vert\phi_{n}^{(j)}(t)\rangle$, respectively.  Notice that the wavefunctions and eigenfunctions differ by just a time-dependent complex-phase, which is computed from~\eqref{eq:schr} through the expectation value
\begin{equation}
\frac{d}{dt}\theta_{n}^{(j)}(t)=\langle \phi_{n}^{(j)}(t)\vert \left[ i\partial_t-\hat{H}_{j}(t) \right]\vert\phi_{n}^{(j)}(t)\rangle \, , \quad j=1,2.
\label{eq:dtheta}
\end{equation}
provided that the Hamiltonian is already known. However, in our case, both the Hamiltonian and the complex-phase are unknown, and a workaround should be implemented. To this end, we consider the time-evolution operator $\hat{U}_{j}(t;t_{0})$, that is, an operator that maps a solution defined at a time $t_{0}$ into one defined at a time $t$, $\vert\psi_{n}^{(j)}(t)\rangle=\hat{U}_{j}(t;t_{0})\vert \psi_{n}^{(j)}(t_{0})\rangle$. Given that both the Hamiltonians $\hat{H}_{j}(t)$ and the quantum invariants $\hat{I}_{j}(t)$ are self-adjoint, it follows that the time-evolution operator is unitary and it takes the diagonal form
\begin{equation}
\hat{U}_{j}(t;t_{0}):=\sum_{n=0}^{\infty}\vert\psi_{n}^{(j)}(t)\rangle\langle\psi_{n}^{(j)}(t_{0})\vert=\sum_{n=0}^{\infty}e^{i\left[\theta_{n}^{(j)}(t)-\theta_{n}^{(j)}(t_{0}) \right]}\vert\phi_{n}^{(j)}(t)\rangle\langle\phi_{n}^{(j)}(t_{0})\vert \, ,
\label{eq:time-evol}
\end{equation}
where it is worth to recall that $\langle \phi_{n}^{(j)}(t')\vert\phi_{m}^{(j)}(t)\rangle\neq\delta_{n,m}$, for $t'\neq t$. Therefore, the set $\{ \vert\psi_{n}^{(j)}(t)\rangle \}_{n=0}^{\infty}$ inherits the orthogonality from the set $\{ \vert\phi_{n}^{(j)}(t)\rangle \}_{n=0}^{\infty}$. In this form, we can be built-up the vector spaces $\overline{\mathcal{H}}_{j}(t)=Span\{ \vert\phi_{n}^{(j)}(t)\rangle \}_{n=0}^{\infty}$, which under the definition of inner-product are equivalent to the vector spaces $\mathcal{H}_{j}(t)$ introduced in Sec.~\ref{sec:Painleve}.

Now, substituting~\eqref{eq:time-evol} and $\vert\psi_{n}^{(j)}(t)\rangle=\hat{U}_{j}(t;t_{0})\vert\psi_{n}^{(j)}(t_{0})\rangle$ into~\eqref{eq:schr} lead us to an expression for the Hamiltonian $\hat{H}_{j}(t)$ in terms of $\hat{U}_{j}(t;t_{0})$ as
\begin{equation}
\hat{H}_{j}(t)=\left[i\partial_{t}\hat{U}_{j}(t;t_{0})\right]\hat{U}_{j}^{\dagger}(t;t_{0})=-\sum_{n=0}^{\infty}\dot{\theta}_{n}^{(j)}(t)\vert\phi_{n}^{(j)}(t)\rangle\langle\phi_{n}^{(j)}\vert+i\sum_{n=0}^{\infty}\left[\partial_{t}\vert\phi_{n}^{(j)}(t)\rangle\right]\langle\phi_{n}^{(j)}(t)\vert \, .
\label{eq:Hj}
\end{equation}
Therefore, from~\eqref{eq:Hj}, the Hamiltonian $\hat{H}_{j}(t)$ is determined once the complex-phase $\theta_{n}^{(j)}(t)$ has been specified. Moreover, it is straightforward to show that the Hamiltonian obtained from~\eqref{eq:Hj} is such that $\hat{I}_{j}(t)$ is its respective quantum invariant. Such a conclusion holds true regardless of the choice of $\theta_{n}^{(j)}(t)$. From~\eqref{eq:Hj} we have to point out that $\hat{H}_{j}(t)$ is composed by the sum of a diagonal and a non-diagonal operator. Thus, in general, the $\hat{H}_{j}(t)$ is not diagonizable in $\mathcal{H}_{j}(t)$. Moreover, since $\hat{H}_{j}(t)$ and $\hat{I}_{j}(t)$ do not commute, a common basis that simultaneously diagonalizes both operators does not exist. 

Now, the time-dependent Hamiltonians related to the quantum invariants $\hat{I}_{j}$ of Sec.~\ref{sec:Painleve} are determined by proposing $\hat{H}_{j}(t)$ as the sum of a kinetic energy term and a time-dependent potential energy term $V_{j}(x,t)$. Given that the complex-phase is arbitrary, we introduce it in the convenient form 
\begin{equation}
\dot{\theta}_{n}^{(j)}(t)\equiv\frac{d}{dt}\theta_{n}^{(j)}(t)=-\frac{\Lambda_{n}^{(j)}}{\sigma^{2}(t)} \, ,
\label{eq:theta}
\end{equation}
leading to the time-dependent Hamiltonians
\begin{equation}
\hat{H}_{j}(t)=\frac{1}{\sigma^{2}}\hat{I}_{j}(t)+\hat{F}(t) \, , \quad \hat{F}(t):=i\sum_{n=0}^{\infty}\left[\partial_{t}\vert\phi_{n}^{(j)}(t)\rangle\right]\langle\phi_{n}^{(j)}(t)\vert \, ,
\label{eq:Hjsimplified}
\end{equation}
where the first part of the Hamiltonian becomes proportional to the invariant operator $\hat{I}_{j}(t)$, and the factor $\sigma^{-2}(t)$ has been introduced such that we recover the kinetic energy term $\hat{p}^{2}$. The operator $\hat{F}(t)$ can be simplified by using the coordinate representation $\langle x \vert \partial_{t}\vert\phi_{n}^{(j)}\rangle$, where the nonstationary eigenfunctions obtained in Sec.~\ref{sec:specI1I2} are all of the form $\phi_{n}^{(j)}(x,t)=e^{i\dot{\sigma}x^{2}/4\sigma}\sigma^{-1/2}K(z(x,t))$, with $z(x,t)=x/\sigma$, and $K(z(x,t))$ a function that depends explicitly on $z$, and implicitly on $x$ and $t$. After some calculations we obtain 
\begin{equation}
\partial_{t}\vert\phi_{n}^{(j)}(t)\rangle=\left[ \frac{i}{4}\left(\frac{\ddot{\sigma}}{\sigma}+\frac{\dot{\sigma}^{2}}{\sigma^{2}} \right)\hat{x}^{2}-\frac{i}{2}\frac{\dot{\sigma}}{\sigma}\{\hat{x},\hat{p} \} \right]\vert\phi_{n}^{(j)}\rangle \, .
\label{eq:partialphi}
\end{equation}
From the latter result, together with the Ermakov equation~\eqref{eq:ermakov}, the time-dependent Hamiltonians take the final form
\begin{equation}
\hat{H}_{j}(t)=\hat{p}^{2}+V_{j}(\hat{x},t) \, , \quad V_{j}(x,t)=\Omega^{2}(t)x^{2}+\frac{1}{\sigma^{2}}R_{j}(x,t) \, , \quad j=1,2,
\label{eq:Hjfinal}
\end{equation}
where $\hat{x}\equiv x$ and $\hat{p}\equiv -i\partial_{x}$ stand for the position and momentum operators, respectively, and the time-dependent potentials $V_{j}(x,t)$ are written in terms of the functions $R_{j}(x,t)$ given in~\eqref{eq:w2}. Notice that a different choice of $\theta_{n}^{(j)}(t)$ lead to a Hamiltonian that, in general, can not be written in terms of the position and momentum operators. The physical meaning of such Hamiltonians is not clear, and will not be considered in the rest of this work.

Finally, the solutions to the Schr\"odinger equation $\psi_{n}^{(j)}(x,t)$ given in~\eqref{eq:schr} are simplified, by using the solutions of the Ermakov equation~\cite{Bla18,Zel19,Zel19b}, as
\begin{equation}
\begin{aligned}
& \psi_{n}^{(j)}(x,t)=e^{i\theta^{(j)}_{n}(t)}\phi_{n}^{(j)}(x,t) \, , \\
& \theta_{n}^{(j)}(t)=-\Lambda_{n}^{(j)}\int^{t}\frac{dt'}{\sigma^{2}(t')}=-\frac{\Lambda_{n}^{(j)}}{2}\arctan\left[\frac{W_{0}}{2}\left(\sqrt{a c-\frac{4}{W_{0}^{2}}}+c\frac{q_{1}(t)}{q_{2}(t)} \right) \right] \, ,
\end{aligned}
\label{eq:psifinal}
\end{equation}
with $a,c$ some arbitrary positive constants given in~\eqref{eq:nonlinear}, and $W_{0}$ the Wronskian of two linearly independent solutions $q_{1,2}(t)$ of the linear equation~\eqref{eq:lineal}. 

We thus have properly identified the time-dependent Hamiltonians whose potential energy term is related to the solutions of the fourth Painlev\'e transcendent. 


\section{Frequency profiles and solutions of the Ermakov equation}
\label{sec:Ermakov}
In this section, we discuss the specific form of $\sigma(t)$ by considering some particular forms of the time-dependent frequency term $\Omega(t)$ that appears in the new time-dependent potentials $V_{j}(x,t)$ given in~\eqref{eq:Hjfinal}, and in the parametric oscillator Hamiltonian~\eqref{eq:H0}. With the solutions to the Ermakov equation properly identified, we determine the reparametrization $z(x,t)=x/\sigma(t)$, and equivalently $y(x,t)=\sqrt{\lambda}z(x,t)$, which are singular-free at each time (see Sec.~\ref{sec:PO}). We have to remark that the form of $V_{j}(x,t)$ depend on the solutions of the Ermakov and fourth Painlev\'e equations. However, the latter are independent one of the other, and thus the solutions constructed in this section are valid for any solution of the fourth Painlev\'e equation discussed in Sec.~\ref{sec:solPainleve}.

To exemplify our results, we consider two different time-dependent frequency profiles. First, we consider the simplest constant frequency case, for which time-dependent potentials are achieved in the most general case, and the stationary results are determined as a particular limit. On the other hand, we consider a frequency profile that changes smoothly from a constant value at $t\rightarrow-\infty$ to another different constant at $t\rightarrow\infty$. Such a profile can be seen as a regularization of the Heaviside distribution~\cite{Olv10}.

\subsubsection*{Frequency $\Omega^{2}(t)=1$}
In this case, two linearly independent solutions of~\eqref{eq:lineal} are given by  $q_{1}(t)=\cos 2(t-t_{0})$ and $q_{2}(t)=\sin 2(t-t_{0})$, with $t_{0}\in\mathbb{R}$ an arbitrary initial time and the Wronskian $W_{0}=2$. After some calculations we obtain 
\begin{equation}
\sigma(t)=\left[\frac{a+c}{2}+\sqrt{a c - 1}\sin 4(t-t_{0})+\frac{a-c}{2}\cos 4(t-t_{0}) \right]^{1/2}
\label{eq:constant1}
\end{equation}
with $a,c>0$ such that $ac\geq 1$. Notice that, even if the frequency $\Omega(t)$ is a constant, the resulting potentials $V_{1,2}(x,t)$ are in general time-dependent, and periodic functions in time. This class of systems are usually studied under the Floquet theory, and already discussed for the parametric oscillator in~\cite{Gla92}.

For $a=c=1$, it follows that $\sigma(t)=1$ and $z(x,t)=x$. Thus, the conventional stationary results reported in~\cite{And00,Mar09} are recovered.

\subsubsection*{Frequency $4\Omega^{2}(t)=\Omega_1+\Omega_2 \tanh(k t)$}
\begin{figure}
\centering
\subfloat[][]{\includegraphics[width=0.4\columnwidth]{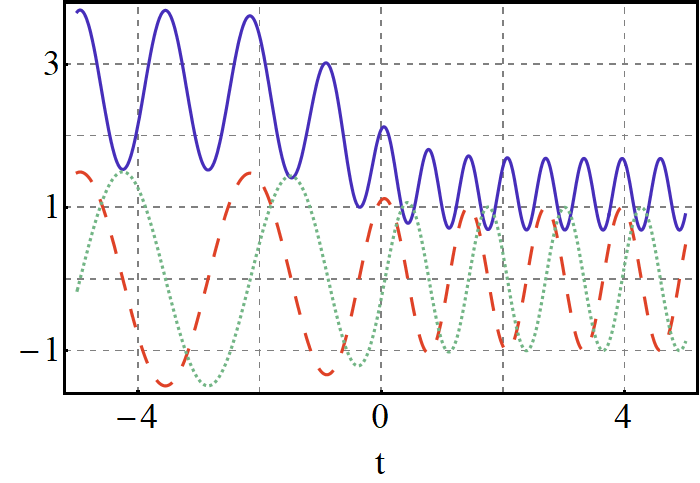}}
\caption{\footnotesize{Functions $\sigma(t)$ (solid-blue), $q_{1}(t)$ (dashed-red), and $q_{2}(t)$ (dotted-green) for the frequency profile $4\Omega^{2}(t)=\Omega_1+\Omega_2 \operatorname{tanh}(k t)$. The parameters has been fixed to $k=1/2$, $\Omega_1=15$, $\Omega_2=10$ and $a=1/2$.}}
\label{fig:SIG1}
\end{figure}

In this case, we introduce the constraint $\Omega_1>\Omega_2$ to ensure that $\Omega(t)$ is a positive function at each time. Exact solutions can be determined for any value of the parameters $k$, $\Omega_{1}$ and $\Omega_{2}$ by taking the linear differential equation~\eqref{eq:lineal} into the hypergeometric form~\cite{Nik88}. After some calculations we obtain two linearly independent solutions
\begin{equation}
\begin{aligned}
& q_{1}(t)=(1-\mathfrak{T}(t))^{-\frac{i}{2}r_{+}}(1+\mathfrak{T}(t))^{-\frac{i}{2}r_{-}} \, {}_{2}F_{1}\left( \left. \begin{aligned} -i \mu \, , \, 1-i \mu \\ 1-ir_{+} \hspace{5mm} \end{aligned} \right\vert \frac{1-\mathfrak{T}(t)}{2} \right) \, , \\
& q_{2}(t)=[q_{1}(t)]^{*} \, , \quad  \mu=\frac{1}{k}\sqrt{\frac{\Omega_1+\sqrt{\Omega_1^{2}-\Omega_2^{2}}}{2}} \, , \quad r_{\pm}=\mu \pm\frac{\Omega_2}{2k^{2}\mu} \, , \quad \mathfrak{T}(t)=\tanh(k t) \, ,
\end{aligned}
\label{eq:TDF2}
\end{equation} 
with ${}_{2}F_{1}(a,b;c;z)$ the \textit{hypergeometric function}~\cite{Olv10}. Given that $q_{2}(t)=[q_{1}(t)]^{*}$, it is trivial to realize that the respective Wronskian becomes $W_{0}=q_{1}\dot{q}_{2}-\dot{q}_{1}q_{2}=-2ikr_{+}$, that is, a pure imaginary constant. Thus, a real-valued solution $\sigma(t)$ is determined if $a=c$ in~\eqref{eq:nonlinear}, leading to
\begin{equation}
\sigma^{2}(t)=2a\operatorname{Re}[q_{1}^{2}(t)]+2\sqrt{a^2 +\frac{1}{k^{2}r_{+}^{2}}} \, \vert q_{1}(t) \vert^{2} \, .
\label{eq:TDF3}
\end{equation}
The behavior of $\sigma(t)$ and $q_{1,2}(t)$ is depicted in Fig.~\ref{fig:SIG1}. For asymptotic times $\vert t \vert>>1$, the frequency function $\Omega(t)$ converges to a constant value, and the respective linear solutions $q_{1,2}(t)$ approximate to sinusoidal functions. Thus, the resulting time-dependent potential $V_{1}(x,t)$ behaves as a periodic function in the asymptotic limit.

\section{Solutions of the Painlev\'e equation}
\label{sec:solPainleve}
As discussed in Sec.~\ref{sec:Painleve}, the solutions to the fourth Painlev\'e equation $w(y)$ allow us to construct the functions $R_{j}(x,t)$ required to determine the time-dependent potentials $V_{j}(x,t)$ given in~\eqref{eq:Hjfinal}. Also, the finite-norm condition of the zero-mode solutions discussed Sec.~\ref{sec:specI1I2} depends strongly on the asymptotic behavior and regularity of $w(y)$. The fourth Painlev\'e equation has been widely studied in the literature~\cite{Gro02,Mar06,Bas92}, and in this section we discuss some hierarchies of solutions that can be implemented in the construction of time-dependent systems. To this end, let us recall the fourth Painlev\'e equation,
\begin{equation}
w_{yy}=\frac{(w_{y})^{2}}{2w}+\frac{3}{2}w^{3}+4yw^{2}+2(y^{2}-\alpha)w+\frac{\beta}{w} \, ,
\label{eq:painleve4}
\end{equation}
whose solutions $w\equiv w(y;\alpha,\beta)$ are determined according to the values of the parameters $\alpha$ and $\beta$. From the latter, the time-dependent Hamiltonian $\hat{H}_{1}(t)$ given in~\eqref{eq:w2} is defined in terms of the time-dependent potential
\begin{equation}
V_{1}(x,t)=\left[\Omega^{2}(t)+\frac{\lambda^{2}-1}{\sigma^{4}}\right]x^{2}-\frac{\lambda}{\sigma^{2}}\left[ \partial_{y}w-w^{2}-2yw+1 \right] \, , \quad y=\sqrt{\lambda}z=\sqrt{\lambda}\frac{x}{\sigma(t)} \, ,
\label{eq:potV1}
\end{equation}
and the respective zero-modes are given in Sec.~\ref{sec:specI1I2}. Throughout the rest of this section, we consider three different hierarchies of solutions $w(y;\alpha,\beta)$, namely the Riccati-like, rational, nonlinear solutions. Those hierarchies have been considered because the spectral information obtained in each case reveals the existence of sequences of solutions, which can be either truncated, infinite, or a combination of both. In addition, we obtain eigenvalues that are equidistant, or equidistant with intermediate gaps.

\subsection{Solutions in terms of the Riccati equation}
\label{subsec:riccati}
It is well-known that the solutions of the fourth Painlev\'e equation can be determined through a Riccati equation~\cite{Gro02} of the form
\begin{equation}
w_{y}=\mu w^{2}+2\mu y w-2(1+\alpha\mu) \, , \quad \mu^{2}=1 \, ,
\label{eq:riccati}
\end{equation}
provided that $\beta=-2(1+\alpha\mu)^{2}$, with $\alpha\in\mathbb{C}$. In this form, we have to solve~\eqref{eq:riccati}, which can be linearized with ease~\cite{Inc56} though the use of a logarithmic derivative as
\begin{equation}
w=-\frac{1}{\mu}\frac{u_{y}}{u} \, , \quad u_{yy}-2\mu y u_{y}-2\mu(1+\alpha\mu)u=0 \, .
\label{eq:riclinear}
\end{equation}
For the physical case under consideration, the set of Painlev\'e parameters $\{\alpha,\beta\}$ are related to the set of physical parameters $\{\lambda, \gamma, d\}$ through the relationships given in~\eqref{eq:GREP}. In this form, for the Riccati-like solutions of~\eqref{eq:painleve4}, we obtain the constraint $d=-[\lambda(1+\mu)+\mu\gamma]^{2}$, where the reducible condition $d\leq 0$ of Sec.~\ref{sec:specI1I2} is automatically fulfilled. In general, the linearized equation~\eqref{eq:riclinear} has two linearly independent solutions of the form
\begin{equation}
u_{1}(y)={}_{1}F_{1}\left(\frac{1}{2}+\mu\frac{\lambda+\gamma}{2\lambda} ,\frac{1}{2};\mu y^{2}\right) \, , \quad u_{2}(y)=\sqrt{\mu y^{2}} \, {}_{1}F_{1}\left(1+\mu\frac{\lambda+\gamma}{2\lambda} ,\frac{3}{2};\mu y^{2}\right) \, ,
\label{eq:u1u2mu}
\end{equation}
with ${}_{1}F_{1}(a,b;z)$ the confluent hypergeometric function~\cite{Olv10}. 

Interestingly, from~\eqref{eq:riccati}, one realizes that the time-dependent potential~\eqref{eq:potV1} reduces, for $\mu=1$, to 
\begin{equation}
V_{1}(x,t)=\left[\Omega^{2}(t)+\frac{\lambda^{2}-1}{\sigma^{4}(t)}\right]x^{2}+\frac{\lambda}{\sigma^{2}(t)}\left(3+\frac{\gamma}{\lambda}\right) \, ,
\label{eq:POshape}
\end{equation}
which in the context of time-dependent systems corresponds to a shape-invariant potential of the parametric oscillator (see discussion in Sec.~\ref{sec:Painleve}). The latter holds for any linear combination of the solutions~\eqref{eq:u1u2mu}. We thus discard the case $\mu=1$ for the rest of the text.

In turn, for $\mu=-1$, the new time-dependent potential takes the form
\begin{equation}
V_{1}(x,t)=\left[\Omega^{2}(t)+\frac{\lambda^{2}-1}{\sigma^{4}(t)}\right]x^{2}-\frac{\lambda}{\sigma^{2}(t)}\left[2\partial_{y}w-2\frac{\gamma}{\lambda}+1\right] \, ,
\label{eq:POnonshape}
\end{equation}
where now we have, in general, a potential different from the class of shape-invariants. From~\eqref{eq:u1u2mu}, we can either choose
\begin{equation}
u_{1}(y)={}_{1}F_{1}\left(-\frac{\gamma}{2\lambda} ,\frac{1}{2};-y^{2}\right)\, , \quad u_{2}(y)=iy \, {}_{1}F_{1}\left(\frac{\lambda-\gamma}{2\lambda} ,\frac{3}{2};-y^{2}\right) \, ,
\label{eq:u1u2}
\end{equation}
or the equivalent \textit{Kummer transformations}~\cite{Olv10}
\begin{equation}
u_{1}(y)=e^{-y^{2}} {}_{1}F_{1}\left(\frac{\gamma+\lambda}{2\lambda} ,\frac{1}{2};y^{2}\right) \, , \quad u_{2}(y)=y \, e^{-y^{2}} {}_{1}F_{1}\left(\frac{2\lambda+\gamma}{2\lambda} ,\frac{3}{2};y^{2}\right) \, .
\label{eq:u1u2-2}
\end{equation}
as the set of linearly independent solutions. For~\eqref{eq:u1u2}, the general solution is constructed as the linear combination 
\begin{equation}
u(y)=k_{a}u_{1}(y)+k_{b}u_{2}(y), \quad \left\vert\frac{k_{a}}{k_{b}}\right\vert>\frac{\Gamma\left(\frac{3}{2}\right)\Gamma\left(\frac{\lambda+\gamma}{2\lambda}\right)}{\Gamma\left(\frac{1}{2}\right)\Gamma\left(\frac{2\lambda+\gamma}{2\lambda}\right)} \, ,
\label{eq:GenLin}
\end{equation}
where the imaginary number $i$ has been absorbed in the constant $k_{b}$, and constraint between the real constants $k_{a}$ and $k_{b}$ is determined from the asymptotic behavior of the confluent hypergeometric function to ensure the existence of a nodeless solution $u(y)$ for $y\in\mathbb{R}$. The latter is required to avoid singularities in the solution of the fourth Painlev\'e equation given in~\eqref{eq:riclinear}, and consequently the potential $V_{1}(x,t)$ in~\eqref{eq:POnonshape}. Similar results are obtained by using the solutions~\eqref{eq:u1u2-2} instead. 

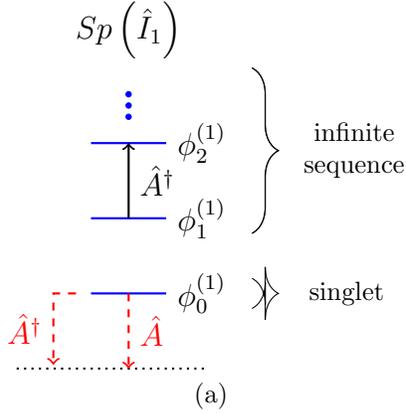
\begin{figure}
\centering
\subfloat[][]{\begin{tikzpicture}
\draw (0.5,2.5) node {$Sp\left(\hat{I}_{1}\right)$};
\filldraw[blue] (0.5,1.35) circle (1pt);
\filldraw[blue] (0.5,1.5) circle (1pt);
\filldraw[blue] (0.5,1.65) circle (1pt);
\draw[blue,thick] (1,1)--(0,1) node[at start,anchor=west,text=black] {$\phi_{2}^{(1)}$};
\draw[red,thick,dashed,->] (-0.2,-1)--(-0.5,-1)--(-0.5,-1.95) node[midway,anchor=east,text=red] {$\hat{A}^{\dagger}$};
\draw[blue,thick] (1,0)--(0,0) node[at start,anchor=west,text=black] {$\phi_{1}^{(1)}$};
\draw[blue,thick] (1,-1)--(0,-1) node[at start,anchor=west,text=black] {$\phi_{0}^{(1)}$};
\draw[black,thick,dotted] (1.5,-2)--(-1,-2);
\draw[thick,black,->] (0.5,0)--(0.5,1) node[midway,anchor=west] {$\hat{A}^{\dagger}$};
\draw[thick,red,dashed,->] (0.5,-1)--(0.5,-2) node[midway,anchor=west] {$\hat{A}$};
\draw [decorate,decoration={brace,amplitude=10pt,raise=4pt},yshift=0pt]
(2,2) -- (2,-0.2) node [black,midway,xshift=1.5cm,yshift=0.25cm] {\footnotesize{infinite}};
\draw [decorate,decoration={brace,amplitude=10pt,raise=4pt},yshift=0pt]
(2,2) -- (2,-0.2) node [black,midway,xshift=1.5cm,yshift=-0.25cm] {\footnotesize{sequence}};
\draw [decorate,decoration={brace,mirror,amplitude=10pt,raise=4pt},yshift=0pt]
(2,-1.2) -- (2,-0.8) node [black,midway,xshift=1.4cm] {\footnotesize{singlet}};
\end{tikzpicture}
\label{fig:Friccati}}
\caption{\footnotesize{(Color online) Ladder structure of the finite-norm zero-modes for the Riccati hierarchy~$\phi_{n}^{(1)}$ obtained in~\eqref{eq:seq-riccati}. The dotted-black line represents the null state, the dashed-red arrow depict the solutions annihilated by either the creation or annihilation operators, and the solid-black arrow represents the transition to higher modes due to the action of $\hat{A}^{\dagger}$.}}
\label{fig:Fseq1}
\end{figure}

Additionally, the asymptotic behavior of the confluent hypergeometric function reveals that $u(y)\rightarrow\infty$ and $w(y)\rightarrow 0$ at $y\rightarrow\pm\infty$. We thus determine the finite-norm elements in $\mathcal{K}_{A}$ and $\mathcal{K}_{A^{\dagger}}$ from~\eqref{eq:ground} and~\eqref{eq:kernelAD}, respectively, leading to the following spectral information
\begin{equation}
\begin{aligned}
& \phi_{0}^{(1)}(x,t)\equiv\phi_{0;1}^{(1)}(x,t)=\Phi_{0;1}^{(1)}(x,t)=\mathcal{N}_{0}^{(1)}\frac{e^{\frac{i}{4}\frac{\dot{\sigma}}{\sigma}x^{2}}}{\sqrt{\sigma}}\frac{e^{-y^{2}/2}}{u(y)} \, , \quad \Lambda_{0}^{(1)}=0 \, , \\
& \phi_{1}^{(1)}(x,t)\equiv\phi_{0;2}^{(2)}(x,t)=\mathcal{N}_{1}^{(1)}\frac{e^{\frac{i}{4}\frac{\dot{\sigma}}{\sigma}x^{2}}}{\sqrt{\sigma}}\left( \frac{u_{y}}{u}+2y\right)e^{-y^{2}/2} \, , \quad \Lambda_{1}^{(1)}=2(\gamma+\lambda) \, ,
\end{aligned}
\label{eq:seq-riccati}
\end{equation}
with $y(x,t)=\sqrt{\lambda}x/\sigma(t)$. From the latter, the zero-mode solution $\phi_{0}^{(1)}$ belongs to both $\mathcal{K}_{A}$ and $\mathcal{K}_{A^{\dagger}}$, that is, $\phi_{0}^{(1)}$ is annihilated by both the creation and annihilation operators. In turn, $\phi_{1}^{(1)}$ is annihilated only by $\hat{A}(t)$, and from it we generate a single sequence of states $\{\phi_{n+1}^{(1)}\}_{n=0}^{\infty}$ through the iterated operation $\phi_{n+1}^{(1)}\propto[\hat{A}^{\dagger}(t)]^{n}\phi_{1}^{(1)}$, up to a normalization constant, with $n=0,1,\cdots$. On the other hand, the respective eigenvalues are determined by increasing $\Lambda_{1}^{(1)}$ in $2\lambda$ units for each iteration, leading to $\Lambda_{n+1}^{(1)}=2[\gamma+\lambda(n+1)]$. For a summary of the spectral information of the quantum invariant, see Fig.~\ref{fig:Fseq1}.

Several special cases can be discussed from the general solution~\eqref{eq:GenLin}, leading to specific hierarchies of solutions of the Painlev\'e equation.

$\bullet$ For $\gamma=0$ together with $2\sqrt{w}k_{a}=1-\sqrt{2\pi}k^{2}$ and $k_{b}=k^{2}$, we obtain the set of parameters $\{\alpha,\beta\}=\{1,0\}$. In such a case we recover the \textit{complementary error function hierarchy} solutions of the form
\begin{equation}
w(y;1,0)=\frac{2\sqrt{2}k^{2}e^{-y^{2}}}{1-\sqrt{2\pi}k^{2}\textnormal{Erfc}(y)} \, ,
\label{eq:CerrorF}
\end{equation}
leading to the equidistant eigenvalues $\Lambda_{n}^{(1)}=2n\lambda $. The respective potential $V_{1}(x,t)$, determined from~\eqref{eq:POnonshape}, reduces to a time-dependent variation of the stationary deformed oscillator potentials reported~\cite{Mie84}. Such a time-dependent potential has been obtained previously in~\cite{Zel17} through the Bagrov-Samsonov approach~\cite{Bag95,Bag96}.

$\bullet$ Another interesting case is recovered for $k_{b}=0$ and $\gamma=2N\lambda$, with $N=0,1,\cdots$, where we obtain the rational solutions
\begin{equation}
u(y)=\mathcal{H}_{2N}(y) \, , \quad w(y;2N+1,-2(2N)^{2})=2N \frac{\mathcal{H}_{2N-1}(y)}{\mathcal{H}_{2N}(y)} \, ,
\label{eq:pseudoHer}
\end{equation}
with $\mathcal{H}_{N}(y)=(-i)^{N}\mathtt{H}_{N}(iy)$ and $\mathtt{H}_{n}(z)$ the pseudo-Hermite and Hermite polynomials~\cite{Olv10}, respectively. Contrary to the previous case, the eigenvalues are non-equidistant and given by $\Lambda_{0}^{(1)}=0$ and $\Lambda_{n+1}^{(1)}=2\lambda(2N+n+1)$. It is worth to remark that the even pseudo-Hermite polynomials are nodeless, whereas the odd ones have one zero at the origin. Thus, the Painlev\'e solution in~\eqref{eq:pseudoHer} is well defined for every $y\in\mathbb{R}$. Such a property is essential since it leads to a rational, nonsingular, and time-dependent potential $V_{1}(x,t)$, where $y\equiv y(x,t)=\sqrt{\lambda}x/\sigma(t)$. This particular case leads to eigenfunctions written in terms of the \textit{exceptional Hermite polynomials}, previously discussed for stationary~\cite{Mar13,Gom14} and time-dependent systems~\cite{Zel19b}.

\subsection{Hierarchies of rational solutions}
\label{subsec:rational}

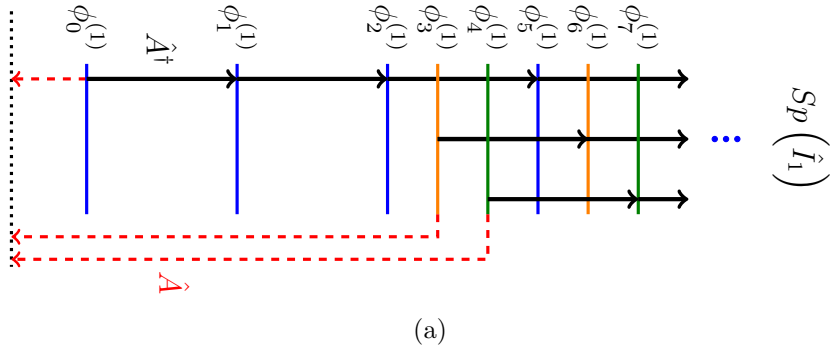
\begin{figure}
\centering
\subfloat[][]{\begin{tikzpicture}
\draw (9.5,1) node[rotate=-90] {$Sp\left(\hat{I}_{1}\right)$};
\filldraw[blue] (8.35,1) circle (1pt);
\filldraw[blue] (8.5,1) circle (1pt);
\filldraw[blue] (8.65,1) circle (1pt);
\draw[black,dotted,very thick] (-1,2.7)--(-1,-0.7);
\draw[red,dashed,very thick,->] (0,1.8)--(-1,1.8);
\draw[red,dashed,very thick,->] (14/3,0)--(14/3,-0.3)--(-1,-0.3);
\draw[red,dashed,very thick,->] (16/3,0)--(16/3,-0.6)--(-1,-0.6) node[midway,anchor=west,text=red,rotate=-90,yshift=-1cm] {$\hat{A}$};
\draw[blue,very thick] (0,2)--(0,0) node[at start,anchor=east,text=black,rotate=-90] {$\phi_{0}^{(1)}$};
\draw[blue,very thick] (6/3,2)--(6/3,0) node[at start,anchor=east,text=black,rotate=-90] {$\phi_{1}^{(1)}$};
\draw[blue,very thick] (12/3,2)--(12/3,0) node[at start,anchor=east,text=black,rotate=-90] {$\phi_{2}^{(1)}$};
\draw[blue,very thick] (18/3,2)--(18/3,0) node[at start,anchor=east,text=black,rotate=-90] {$\phi_{5}^{(1)}$};
\draw[ultra thick,black,->] (0,1.8)--(2,1.8) node[midway,anchor=east,rotate=-90] {$\hat{A}^{\dagger}$};
\draw[ultra thick,black,->] (2,1.8)--(4,1.8);
\draw[ultra thick,black,->] (4,1.8)--(6,1.8);
\draw[ultra thick,black,->] (6,1.8)--(8,1.8);
\draw[orange,very thick] (14/3,2)--(14/3,0) node[at start,anchor=east,text=black,rotate=-90] {$\phi_{3}^{(1)}$};
\draw[orange,very thick] (20/3,2)--(20/3,0) node[at start,anchor=east,text=black,rotate=-90] {$\phi_{6}^{(1)}$};
\draw[ultra thick,black,->] (14/3,1)--(20/3,1);
\draw[ultra thick,black,->] (14/3,1)--(8,1);
\draw[color=green!50!black,very thick] (16/3,2)--(16/3,0) node[at start,anchor=east,text=black,rotate=-90] {$\phi_{4}^{(1)}$};
\draw[color=green!50!black,very thick] (22/3,2)--(22/3,0) node[at start,anchor=east,text=black,rotate=-90] {$\phi_{7}^{(1)}$};
\draw[ultra thick,black,->] (16/3,0.2)--(22/3,0.2);
\draw[ultra thick,black,->] (22/3,0.2)--(8,0.2);
\end{tikzpicture}
\label{fig:Fokamoto}}
\caption{\footnotesize{(Color online) Ladder structure of the finite-norm zero-modes for the ``$-2y/3$'' hierarchy~$\phi_{n}^{(1)}$ obtained in~\eqref{eq:okamoto-zeromodes}. The dotted-black line represents the null state, the dashed-red arrow depict the solutions annihilated by either the creation or annihilation operators, and the solid-black arrow represents the transition to higher modes due to the action of $\hat{A}^{\dagger}$.}}
\label{fig:Fseq2}
\end{figure}

\begin{sidewaysfigure}
\centering
\subfloat[][]{\includegraphics[width=0.27\textwidth]{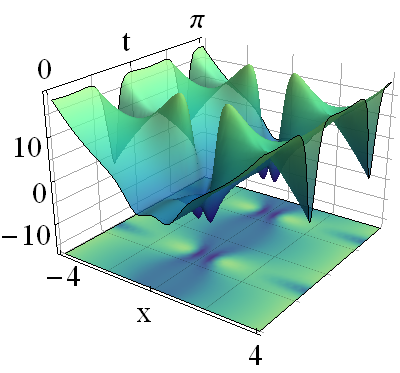}
\label{fig:Oka-pot}}
\hspace{5mm}
\subfloat[][]{\includegraphics[width=0.2\textwidth]{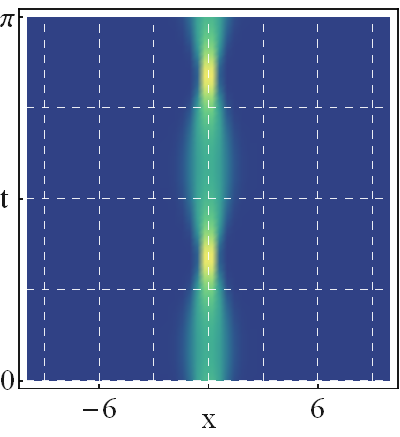}
\label{fig:Oka-wf1}}
\subfloat[][]{\includegraphics[width=0.2\textwidth]{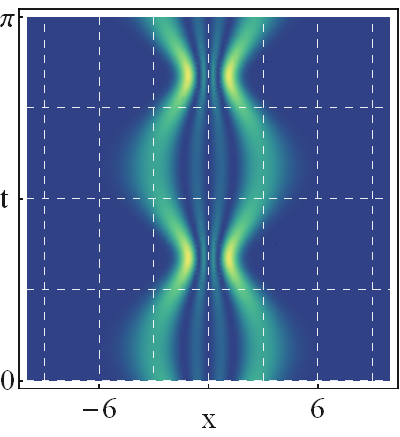}
\label{fig:Oka-wf2}}
\subfloat[][]{\includegraphics[width=0.2\textwidth]{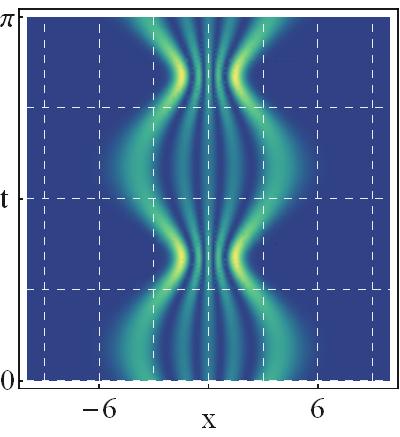}
\label{fig:Oka-wf3}}
\\
\subfloat[][]{\includegraphics[width=0.27\textwidth]{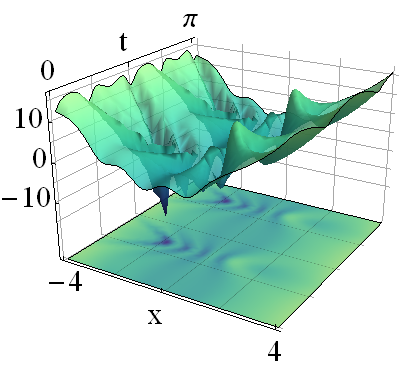}
\label{fig:NL-pot}}
\hspace{5mm}
\subfloat[][]{\includegraphics[width=0.2\textwidth]{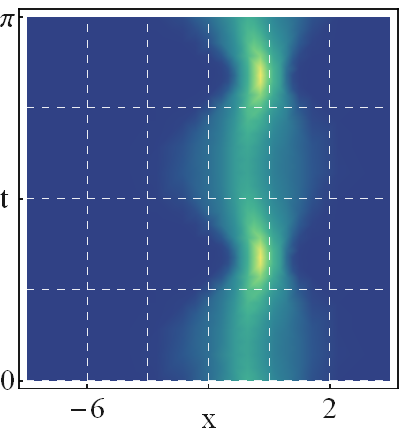}
\label{fig:NL-wf1}}
\subfloat[][]{\includegraphics[width=0.2\textwidth]{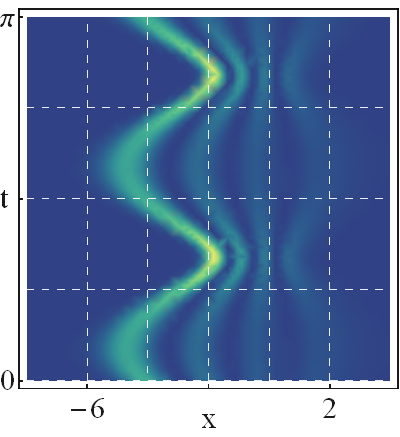}
\label{fig:NL-wf2}}
\subfloat[][]{\includegraphics[width=0.2\textwidth]{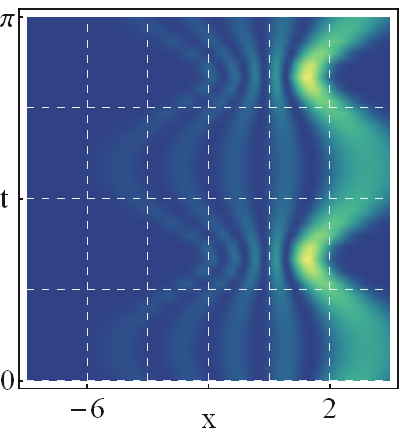}
\label{fig:NL-wf3}}
\caption{(Upper row) Time-dependent potential $V_{1}(x,t)$ (a) computed through the hierarchy of rational solutions $w_{2}$ in terms of the Okamoto polynomials~\eqref{eq:w2-Oka}, together with the probability densities $\vert\phi^{(1)}_{0}\vert^{2}$ (b), $\vert\phi^{(1)}_{3}\vert^{2}$ (c), and $\vert\phi^{(1)}_{4}\vert^{2}$ (d) of the zero-modes given in~\eqref{eq:okamoto-zeromodes}. The frequency profile is $\Omega^{2}(t)=1$, and the parameters have been fixed to $\lambda=2.5$, $a=2$, and $c=1$.\\
(Lower row) Time-dependent potential $V_{1}(x,t)$ (e) computed through the hierarchy of nonlinear bound states $w=2\sqrt{2}\eta^{2}_{k}(\chi,N)$ of Sec.~\ref{subsec:nonlinb}, together with the probability densities $\vert\phi^{(1)}_{0}\vert^{2}$ (f), $\vert\phi^{(1)}_{N}\vert^{2}$ (g), and $\vert\phi^{(1)}_{N+1}\vert^{2}$ (h) of the zero-modes given in~\eqref{eq:nonlinZero}. The frequency profile is $\Omega^{2}(t)=1$, and the parameters have been fixed to $N=3$, $k=0.44/\sqrt{3!}$, $\lambda=1$, $a=2$, and $c=1$.}
\label{fig:pots}
\end{sidewaysfigure}

In general, it is well-known that the fourth Painlev\'e equation admits hierarchies of rational solutions \textit{if and only if}~\cite{Gro02} the set of parameters $\{\alpha,\beta\}$ take either the values $\{ m,-2(2n-m+1)^{2}\}$ or $\{ m,-2(2n-m+1/3)^{2}\}$. The class of all the rational solutions are classified as ``$-1/y$'', ``$-2y$'', and ``$-2y/3$'' hierarchies. For instance, it has been shown that solutions in terms of the \textit{generalized Hermite polynomials} $H_{n,m}(z)$ contain all solutions in the ``$-1/y$'' and ``$-2y$'' hierarchies, whereas the \textit{generalized Okamoto polynomials} $Q_{n,m}(y)$ determine the ``$-2y/3$'' hierarchy. For a complete discussion see~\cite{Gro02}. The relation of such rational solutions with SUSYQM in the stationary regime has been already discussed in~\cite{Mar16}. For the sake of simplicity, and to illustrate our general set-up, let us consider the simplest family of solutions in the ``$-2y/3$'' hierarchy, that is,
\begin{equation}
w_{M}\equiv w(y;2M,-2/9)=-\frac{2y}{3}+\frac{\partial}{\partial y}\ln\left(\frac{Q_{M+1}}{Q_{M}}\right) \, ,
\label{eq:rat1}
\end{equation}
where $Q_{M}\equiv Q_{M}(y)$ stands for the \textit{Okamoto polynomials}~\cite{Oka86}. The latter are determined from the nonlinear recurrence relationship
\begin{equation}
Q_{M+1}=-\frac{9}{2}\left(\frac{Q_{M}Q''_{M}-[Q_{M}']^{2}}{Q_{M-1}} \right)+\left[2y^2+3(2M-1) \right]\frac{Q_{M}^{2}}{Q_{M-1}} \, , \quad M=1,2,\cdots ,
\label{eq:okamoto-rec}
\end{equation}
with $Q_{0}=Q_{1}=1$ and $Q_{M}'\equiv\partial Q_{M}/\partial y$. Given that the Okamoto polynomials do not contain zeros for real $y$, one conclude that $w_{M}$ given in~\eqref{eq:rat1} is a singular-free solution. To simplify our calculations, let us consider $M=2$, in such a case we obtain 
\begin{equation}
w_{2}=-\frac{2y}{3}+\frac{16 y^3 \left(4y^4+24y^2+45\right)}{\left(2 y^2+3\right) \left(8 y^6+60 y^4+90y^2+135\right)} \, ,
\label{eq:w2-Oka}
\end{equation}
together with the finite-norm zero modes
\begin{equation}
\begin{aligned}
&\phi_{0}^{(1)}(x,t)\equiv\phi_{0;1}^{(1)}=\mathcal{N}_{0}^{(1)}\frac{e^{\frac{i}{4}\frac{\dot{\sigma}}{\sigma}x^{2}}}{\sqrt{\sigma}}e^{-\frac{y^2}{6}}\frac{\left(2y^2+3\right)}{8y^6+60y^4+90y^2+135} \, , \\
&\phi_{3}^{(1)}(x,t)\equiv\phi_{0;2}^{(1)}=\mathcal{N}_{3}^{(1)}\frac{e^{\frac{i}{4}\frac{\dot{\sigma}}{\sigma}x^{2}}}{\sqrt{\sigma}} e^{-\frac{y^2}{6}}\frac{y\left(8 y^4 \left(2y^4+24 y^2+63\right)-2835\right)}{\left(8y^6+60y^4+90y^2+135\right)} \, , \\
&\phi_{4}^{(1)}(x,t)\equiv\phi_{0;3}^{(1)}=\mathcal{N}^{(1)}_{4}\frac{e^{\frac{i}{4}\frac{\dot{\sigma}}{\sigma}x^{2}}}{\sqrt{\sigma}}e^{-\frac{y^2}{6}}\frac{\left(2 \left(4y^2 \left(2y^2+15\right) \left(2y^4-45\right)-6075\right) y^2+6075\right)}{\left(8y^6+60 y^4+90y^2+135\right)} \, ,
\end{aligned}
\label{eq:okamoto-zeromodes}
\end{equation}
where $\mathcal{N}_{n}^{(1)}$ stands for the respective normalization constants. From~\eqref{eq:okamoto-zeromodes} one can see that $\phi_{0}^{(1)}$, $\phi_{3}^{(1)}$, and $\phi_{4}^{(1)}$ have exactly zero, three, and four real nodes, respectively. Moreover, the associated eigenvalues are $\Lambda_{0}^{(1)}=0$, $\Lambda_{3}^{(1)}=14\lambda/3$, and $\Lambda_{4}^{(1)}=16\lambda/3$. Every zero-mode in~\eqref{eq:okamoto-zeromodes} is an element of $\mathcal{K}_{\hat{A}}$, and consequently, each mode generates an infinite sequence of solutions, that is, we have three infinite sequences. The behavior of the respective potential and the probability densities associated with the zero-modes is depicted in Figs.~\ref{fig:Oka-pot}-\ref{fig:Oka-wf3}, where the frequency profile has been fixed as a constant, $\Omega^{2}(t)=1$, with $\sigma(t)$ given in~\eqref{eq:constant1}. Notice that, even in such a case, the resulting potential depends explicitly on time.

\subsection{Solutions in terms of nonlinear bound states}
\label{subsec:nonlinb}
\begin{figure}
\centering
\subfloat[][]{\includegraphics[width=0.45\textwidth]{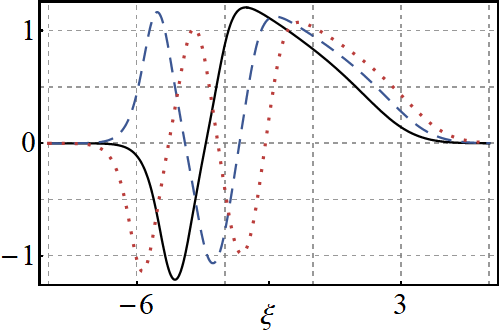}
\label{fig:nonlin-plot}}
\hspace{5mm}
\subfloat[][]{\begin{tikzpicture}
\draw (0.5,4.5) node {$Sp\left(\hat{I}_{1}\right)$};
\filldraw[blue] (0.5,3.35) circle (1pt);
\filldraw[blue] (0.5,3.5) circle (1pt);
\filldraw[blue] (0.5,3.65) circle (1pt);
\draw[blue,thick] (1,3)--(0,3) node[at start,anchor=west,text=black] {$\phi_{N+2}^{(1)}$};
\draw[black,thick,->] (0.5,2)--(0.5,3) node[midway,anchor=west,text=black] {$A^{\dagger}$};
\draw[blue,thick] (1,2)--(0,2) node[at start,anchor=west,text=black] {$\phi_{N+1}^{(1)}$};
\draw[red,thick,dashed,->] (-0.2,2)--(-0.75,2)--(-0.75,-1.95) node[midway,anchor=east,dashed,text=red] {$A$};
\draw[blue,thick] (1,1)--(0,1) node[at start,anchor=west,text=black] {$\phi_{N}^{(1)}$};
\draw[red,thick,dashed,->] (-0.2,1)--(-0.5,1)--(-0.5,-1.95) node[midway,anchor=west,text=red] {$A^{\dagger}$};
\draw[blue,thick] (1,0)--(0,0) node[at start,anchor=west,text=black] {$\phi_{1}^{(1)}$};
\draw[blue,thick] (1,-1)--(0,-1) node[at start,anchor=west,text=black] {$\phi_{0}^{(1)}$};
\draw[black,thick,dotted] (1.5,-2)--(-1,-2);
\draw[thick,black,->] (0.5,-1)--(0.5,0) node[midway,anchor=west] {$A^{\dagger}$};
\draw[thick,red,dashed,->] (0.5,-1)--(0.5,-2) node[midway,anchor=west] {$A$};
\filldraw[blue] (0.5,0.35) circle (1pt);
\filldraw[blue] (0.5,0.5) circle (1pt);
\filldraw[blue] (0.5,0.65) circle (1pt);
\draw [decorate,decoration={brace,amplitude=10pt,raise=4pt},yshift=0pt]
(2,1.3) -- (2,-1.3) node [black,midway,xshift=1.7cm,yshift=0.25cm] {\footnotesize{$(N+1)-$dim.}};
\draw [decorate,decoration={brace,amplitude=10pt,raise=4pt},yshift=0pt]
(2,1.3) -- (2,-1.3) node [black,midway,xshift=1.7cm,yshift=-0.25cm] {\footnotesize{sequence}};
\draw [decorate,decoration={brace,amplitude=10pt,raise=4pt},yshift=0pt]
(2,4) -- (2,1.7) node [black,midway,xshift=1.5cm,yshift=0.25cm] {\footnotesize{infinite}};
\draw [decorate,decoration={brace,amplitude=10pt,raise=4pt},yshift=0pt]
(2,4) -- (2,1.7) node [black,midway,xshift=1.5cm,yshift=-0.25cm] {\footnotesize{sequence}};
\end{tikzpicture}
\label{fig:nonlin-seq}}
\caption{\footnotesize{(a) (Color online) Solutions $\eta_{k}(\xi;N)$ computed through~\eqref{eq:recursion}-\eqref{eq:recursion0} with the parameters $\{ N=1, k=0.446 \}$ (solid-black), $\{ N=2, k=0.446/\sqrt{2!} \}$ (dashed-blue), and $\{ N=3, k=0.446/\sqrt{3!} \}$ (dotted-red). (b) (Color online) Ladder structure of the finite-norm zero-modes for the non-linear bound state hierarchy $\phi_{n}^{(1)}$ obtained in~\eqref{eq:nonlinZero}. The dotted-black line represents the null state, the dashed-red arrow depict the solutions annihilated by either the creation or annihilation operators, and the solid-black arrow represents the transition to higher modes due to the action of $\hat{A}^{\dagger}$.}}
\label{fig:Fnonlin}
\end{figure}
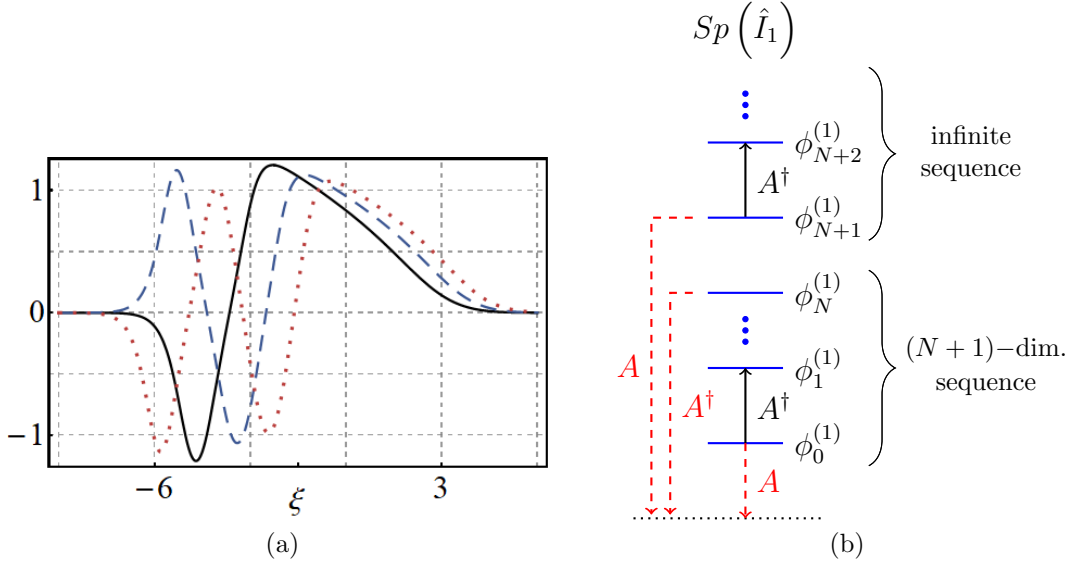

Another special class of solutions to~\eqref{eq:painleve4} is determined by considering the set of parameters $\{\alpha,\beta\}=\{2\nu+1,0\}$, together with the reparametrization 
\begin{equation}
w(y;2\nu+1,0)=2\sqrt{2}\eta_{k}^{2}(\xi;\nu)\, , \quad  y=\frac{\xi}{\sqrt{2}} \, .
\label{eq:nonlin-rep}
\end{equation} 
The latter leads to a nonlinear differential equation for $\eta_{k}(\xi;\nu)$ of the form~\cite{Bas92}	
\begin{equation}
\frac{d^{2}\eta_{k}}{d\xi^{2}}=3\eta_{k}^{5}+2\xi \eta_{k}^{3}+\left(\frac{1}{4}\xi^{2}-\nu-\frac{1}{2}\right)\eta_{k} \, , \quad \eta\equiv \eta_{k}(\xi;\nu) \, , \quad \nu_{k}\in\mathbb{R} \, ,
\label{eq:etak}
\end{equation}
which arises in the study of the \textit{derivative nonlinear Schr\"odinger equation}~\cite{Kau78}. A striking feature of $\eta_{k}$ is provided by the asymptotic behavior $\eta_{k}(\xi;\nu)\sim k D_{\nu}(\xi)$ as $\xi\rightarrow+\infty$ for $\nu\in\mathbb{R}$ and $D_{\nu}(\xi)$ the \textit{parabolic cylinder functions}~\cite{Olv10}. In turn, determining the asymptotic behavior for $\xi\rightarrow-\infty$ becomes a challenging task, where the asymptotic value depends on $\nu$ and it is computed from a connection formulae, see~\cite{Its98} for details. In this section, we restrict ourselves to the special case $\nu=N$, with $N=0,1,\cdots$. In such a case, there are solutions $\eta_{k}(\xi;N)$ for $\xi\in\mathbb{R}$ with asymptotic behavior 
\begin{equation}
\eta_{k}(\xi;N)\sim\begin{cases}k\xi^{N}e^{-\xi^{2}/4} \quad &x\rightarrow+\infty\\ \frac{k\xi^{N}e^{-\xi^{2}/4}}{\sqrt{1-2\sqrt{2\pi}N!k^{2}}} \quad &x\rightarrow-\infty \end{cases} \, , \quad k^{2}<\frac{1}{2\sqrt{2\pi}n!} \, .
\label{eq:asymptotic}
\end{equation}
That is, the solutions decay exponentially to zero at both $\xi\rightarrow\pm\infty$. The exact form of $\eta_{k}(\xi;N)$ is determined in a recursive way through the combination of several B\"acklund transformations such that $\beta=0$ is preserved in each iteration~\cite{Bas92}. We thus have
\begin{equation}
\eta_{\frac{k}{\sqrt{n+1}}}(\xi;N+1)=\frac{\xi\eta_{k}(\xi;N)+2\eta^{3}_{k}(\xi;N)-2\eta'_{k}(\xi;N)}{2\left[N+1+2\eta_{k}(\xi;N)\eta'_{k}(\xi;N)-\xi\eta^{2}_{k}(\xi;N)-2\eta^{4}_{k}(\xi;N)\right]^{1/2}} \, ,
\label{eq:recursion}
\end{equation}
with $\eta'_{k}$ the partial derivative of $\eta_{k}$ with respect to $\xi$. Thus, the $N$-th solution is determined by iterating $N$ times the solution associated with $N=0$ in the recursion formula~\eqref{eq:recursion}. The $N=0$ solution is related to the complementary error function hierarchy~\eqref{eq:CerrorF} as
\begin{equation}
\eta_{k}(\xi;0)\equiv\frac{1}{2^{3/2}}\left[w(\xi/\sqrt{2};1,0)\right]^{1/2}=\frac{ke^{-\xi^{2}/4}}{\left[1-\sqrt{2\pi}k^{2}\textnormal{Erfc}(\xi/\sqrt{2})\right]^{1/2}} \, .
\label{eq:recursion0}
\end{equation}
The behavior of the solutions $\eta_{k}(\xi;N)$ are depicted in Fig.~\ref{fig:nonlin-plot} for several nalues of $N$. In such a figure it can be seen that, indeed, the solutions contain exactly $N$ zeroes while they converge to zero at the boundary points of the domain. 

Now, with the above solutions and their asymptotic behavior, it is straightforward to determine the set of finite-norm zero-mode eigenfunctions. From~\eqref{eq:ground}-\eqref{eq:kernelAD} we obtain
\begin{equation}
\begin{aligned}
&\phi_{0}^{(1)}(x,t)\equiv\phi_{0;1}^{(1)}=\mathcal{N}_{0}^{(1)}\frac{e^{\frac{i}{4}\frac{\dot{\sigma}}{\sigma}x^{2}}}{\sqrt{\sigma}}e^{-\lambda z^{2}/2}e^{-2\int^{z}dz'G(z')} \, , \\
&\phi_{N}^{(1)}(x,t)\equiv\Phi_{0;1}^{(1)}=\Phi_{0;2}^{(1)}=\mathcal{N}_{N}^{(1)}\frac{e^{\frac{i}{4}\frac{\dot{\sigma}}{\sigma}x^{2}}}{\sqrt{\sigma}}e^{-\int^{z}dz'G(z')}\eta_{k}(\xi;N) \, , \\
&\phi_{N+1}^{(1)}(x,t)\equiv\phi_{0;2}^{(1)}=\phi_{0;3}^{(1)}=\mathcal{N}^{(1)}_{N+1}\frac{e^{\frac{i}{4}\frac{\dot{\sigma}}{\sigma}x^{2}}}{\sqrt{\sigma}}e^{\int^{z}dz'G(z')}F_{k}(\xi;N)\eta_{\frac{k}{\sqrt{N+1}}}(\xi;N+1), \\
&F_{k}(\xi;N)=\left[N+1+2\eta_{k}(\xi;N)\eta'_{k}(\xi;N)-\xi\eta^{2}_{k}(\xi;N)-2\eta^{4}_{k}(\xi;N)\right]^{1/2} \, ,
\end{aligned}
\label{eq:nonlinZero}
\end{equation}
with the respective eigenvalues $\Lambda_{0}^{(1)}=0$, $\Lambda_{N}^{(1)}=2\lambda N$, and $\Lambda_{N+1}^{(1)}=2\lambda(N+1)$. From the asymptotic behavior~\eqref{eq:asymptotic}, one realizes that the term $\exp\left(\int^{z}dz'G(z')\right)$ converges to a finite value for $z\rightarrow\pm\infty$, since the integral approximates to the error function at the asymptotic value. Thus, every zero-mode eigenfunction in~\eqref{eq:nonlinZero} converges to zero at $z\rightarrow\pm\infty$ and, indeed, we have finite-norm solutions. The remaining elements of the spectrum are determined from the action of the creation operator $\hat{A}^{\dagger}(t)$ on the zero modes~\eqref{eq:nonlinZero}, as usual. Notice that $\phi_{N}^{(1)}\in\mathcal{K}_{A^{\dagger}}$, and thus $\phi_{N}^{(1)}$ is annihilated by the creation operator $\hat{A}^{\dagger}(t)$. Therefore, the creation operator generates a $(N+1)$-dimensional sequence of eigenfunctions $\{\phi_{n}^{(1)}\}_{n=0}^{N}$ through the iteration $\left[\hat{A}^{\dagger}(t)\right]^{n}\phi_{0}^{(1)}$, for $n=0,1,\cdots,N$. In turn, an additional infinite sequence $\{\phi_{n}^{(1)}\}_{n=N+1}^{\infty}$ is generated from the operation $\left[\hat{A}^{\dagger}(t)\right]^{n}\phi_{N+1}^{(1)}$, for $n=0,1,\cdots$. In this form, the case discussed in this section generalizes the complementary error function hierarchy, where the latter is obtained as the special case $N=0$. This spectral information is summarized in the diagram depicted in Fig.~\ref{fig:nonlin-seq}.

Finally, from the zero modes~\eqref{eq:nonlinZero}, together with the properties of the nonlinear bound states $\eta_{k}(\xi;N)$, it follows that $\phi_{0}^{(1)}$, $\phi_{N}^{(1)}$, and $\phi_{N+1}^{(1)}$ are solutions with exactly zero, $N$, and $N+1$ nodes, respectively. Therefore, the oscillation theorem for the Sturm-Liouville associated with the quantum invariant $\hat{I}_{1}(t)$ is verified. Additionally, given that the action of the creation operator increases the eigenvalue by $2\lambda$ units, we determine that in general the eigenvalues of $\hat{I}_{1}(t)$ are equidistant, $\Lambda_{n}^{(1)}=2\lambda n$, for $n=0,1,\cdots$. The behavior of the respective time-dependent potential $V_{1}(x,t)$ and the probability densities of the zero-modes is depicted in Figs.~\ref{fig:NL-pot}-\ref{fig:NL-wf3}, where we have chosen $\Omega^{2}(t)=1$, with $\sigma(t)$ given in~\eqref{eq:constant1}.


\section{Conclusions}
The results of this manuscript can be seen from two different perspectives. On the one hand our approach represents a time-dependent generalization of the families of potentials reported previously in the stationary regime~\cite{And00}, on the other hand we also introduce some new quantum potentials unnoticed in the literature of stationary models. To this end, it was essential to address the shape-invariant problem from the more general perspective of the quantum invariants rather than the Hamiltonians. In this form, the time-dependence is introduced to both quantities, where the conventional spectral analysis is now carried on for the quantum invariant. Regardless of its time-dependence, the eigenvalues associated with the quantum invariant are time-independent, as it was first proved by Lewis-Reisenfeld~\cite{Lew69}. Interestingly, after introducing the time parameter in the construction, a second nonlinear equation appears, namely the Ermakov equation, in such a way that the resulting time-dependent potentials and solutions to the Schr\"odnger equation are free of singularities at each time. In turn, the fourth Painlev\'e equation emerges after using a convenient reparametrization, where the parameters of the Painelv\'e equation dictate the distribution of eigenvalues of the quantum invariant, provided that the respective zero-modes are physically acceptable. It is worth to mention that both nonlinear equations, Ermakov and Painlev\'e, are not interlaced to each other; that is, the solutions of one equation do not modify the outcome of the solutions of the other equation. We can thus study each equation independently.

Regarding the fourth Painlve\'e equation, a first family of solutions is determined through the related Riccati equation. This does not only allows us to recover the one-step rational extensions of the parametric oscillator, reported previously in~\cite{Zel19b}, but also leads to a family of one-parameter solutions in terms of the error function. The respective potential corresponds to a time-dependent generalization of the deformed oscillator reported by Mielnik~\cite{Mie84}. On the other hand, the hierarchy of rational solutions ``$-2y/3$'' in terms of the Okamoto polynomials allows constructing a quantum invariant with several gaps in its spectrum, which is generated by three infinite sequences of independent eigenfunctions, that is, the respective eigenvalues in each sequence do not overlap. As a particular example, we have shown that the Okamoto polynomial $Q_{2}(y)$ generates two gaps. Nevertheless, our results can be separated for an arbitrary polynomial $Q_{N}(y)$, with $N=0,1,\cdots$, where the spectrum acquires precisely $N$ gaps. The latter is indeed a property that could reveal an intrinsic structure in terms of exceptional polynomials. Further analysis is required, and results on the matter will be reported elsewhere.

Although a closed expression for the $N$-th nonlinear bound state is not available, a nonlinear recurrence relation in the form of a B\"acklund transformation allows computing any solution by iterative means from the seed solution given by the error function. Remarkably, the B\"acklund transformation~\cite{Bas92} is such that preserves $\beta=0$ for any $N$-th nonlinear bound state. The latter implies that the spectrum of the quantum invariant is equidistant, for any $N$. Furthermore, the eigenfunctions are classified by two sequences, one that is $(N+1)$-dimensional and one infinite-dimensional. The finite-dimensional sequence has two zero-modes, constructed as eigenfunctions of the annihilation (nodeless function) and creation ($N$ nodes function) operators. The zero-mode related to the infinite sequence is also an eigenfunction of the annihilation operator such that it has exactly $N+1$ nodes. Interestingly, this case also brings new results in the stationary regime, for it generalizes the singlet and doublet structure introduced in~\cite{And00}. Thus, it is clear that the families of nonlinear bound states can be explored even further in the context of stationary Hamiltonians, and a detailed analysis will be discussed in an upcoming contribution.


\appendix
\setcounter{section}{0}  
\section{Parametric oscillator}
\label{sec:PO}
\renewcommand{\thesection}{A-\arabic{section}}

\renewcommand{\theequation}{A-\arabic{equation}}
\setcounter{equation}{0}  
In this appendix, we briefly introduce the basic notions of the parametric oscillator, also known as \textit{nonstationary oscillator}. The latter system is characterized by a time-dependent quadratic Hamiltonian of the form
\begin{equation}
\hat{H}_{0}(t):=\hat{p}^{2}+\Omega^{2}(t)\hat{x}^{2}\equiv-\frac{\partial^{2}}{\partial x^{2}}+\Omega^{2}(t)x^{2} \, , 
\label{eq:H0}
\end{equation}
with $\hat{x}\equiv x$ and $\hat{p}\equiv -i\partial/\partial x$ the canonical position and momentum operators, respectively, and $\Omega(t)>0$ the time-dependent frequency of oscillation. In contradistinction to the stationary oscillator, the Hamiltonian $\hat{H}_{0}(t)$ does not admit an eigenvalue equation. However, from the approach of Lewis-Riesenfeld~\cite{Lew69}, it is known that solutions of the Schr\"odinger equation 
\begin{equation}
i\frac{\partial}{\partial t}\psi_{n}^{(0)}(x,t)=\hat{H}_{0}(t)\psi_{n}^{(0)}(x,t)\rangle \, ,
\label{eq:schrH0}
\end{equation}
are determined from the eigenvalue problem
\begin{equation}
\hat{I}_{0}(t)\phi_{n}^{(0)}(x,t)=\Lambda^{(0)}_{n}\phi_{n}^{(0)}(x,t) \, ,
\label{eq:EIGENI0}
\end{equation}
with $\Lambda_{n}^{(0)}$ the time-independent eigenvalues and $\phi_{n}^{(0)}(x,t)$ the nonstationary eigenfunctions of the \textit{quantum invariant} $\hat{I}_{0}(t)$ of the system. Such an invariant is computed from the invariance condition
\begin{equation}
\frac{d\hat{I}_{0}(t)}{dt}=i[\hat{H}_{0},\hat{I}_{0}(t)]+\frac{\partial\hat{I}_{0}(t)}{\partial t}=0 \, ,
\label{eq:invcond}
\end{equation}
and it takes the form~\cite{Lew69,Zel19}
\begin{equation}
\hat{I}_{0}(t)=\sigma^{2}\hat{p}^{2}+\left(\frac{\dot{\sigma}^{2}}{4}+\frac{1}{\sigma^{2}} \right)\hat{x}^{2}-\frac{\sigma\dot{\sigma}}{2}\{\hat{x},\hat{p}\} \, , \quad \dot{\sigma}\equiv\frac{d\sigma(t)}{dt} \, ,
\label{eq:invI0}
\end{equation}
with $\sigma=\sigma(t)$ a solution of the nonlinear equation
\begin{equation}
\ddot{\sigma}+4\Omega^{2}(t)\sigma=\frac{4}{\sigma^{3}} \, .
\label{eq:ermakov}
\end{equation}
The latter equation is known as the \textit{Ermakov equation}~\cite{Erm08,Mil30,Pin50}, and a solution is found through the nonlinear combination~\cite{Ros15,Ros18}
\begin{equation}
\sigma(t)=\left[a q_{1}^{2}(t)+b q_{1}(t)q_{2}(t)+cq_{2}^{2}(t) \right]^{1/2} \, , \quad b^{2}-4ac=-\frac{16}{W_{0}^{2}} \, ,
\label{eq:nonlinear}
\end{equation}
with $W_{0}=W(q_{1},q_{2})$ the Wronskian of two linearly independent solutions $q_{1,2}(t)$ of the linear homogeneous equation
\begin{equation}
\ddot{q}_{1,2}+4\Omega^{2}(t)q_{1,2}=0 \, .
\label{eq:lineal}
\end{equation}
From the form of the differential equation~\eqref{eq:lineal}, it is straightforward to realize that the Wronskian $W_{0}$ is time-independent, regardless of the structure of $\Omega^{2}(t)$. The constraint in the constants given in~\eqref{eq:nonlinear} guarantees that $\sigma(t)$ is a nodeless function at any time. Such a feature is essential to construct regular solutions $\psi_{n}^{(0)}(x,t)$, and also in determining new nonsigular time-dependent potentials, as discussed in Sec.~\ref{sec:Painleve}.

The spectral problem~\eqref{eq:EIGENI0} has been already determined in the literature through several techniques, such as solving the differential equation directly~\cite{Lew69}, using a particular complex reparametrization~\cite{Gla92}, with aid of the Fourier transform~\cite{Ram18}, and performing geometrical transformations~\cite{Zel19}. Thus, the spectral information of $\hat{I}_{0}(t)$ is given by
\begin{equation}
\phi_{n}^{(0)}(x,t)=\frac{e^{\frac{i}{2}\frac{\dot{\sigma}}{\sigma}x^{2}}}{\sqrt{2^{n}n!\sigma\sqrt{\pi}}}e^{-\frac{x^{2}}{2\sigma^{2}}}\texttt{H}_{n}\left( \frac{x}{\sigma}\right) \, , \quad \Lambda_{n}^{(0)}=2n+1 \, , \quad n=0,1,\cdots,
\label{eq:solutionsI0}
\end{equation}
with $\texttt{H}_{n}(z)$ the Hermite polynomials~\cite{Olv10}. Clearly, the elements of the set $\{ \phi_{n}^{(0)} \}_{n=0}^{\infty}$ do not fulfill~\eqref{eq:schrH0}, but it can be easily shown that 
\begin{equation}
\psi_{n}^{(0)}(x,t)=e^{i\theta^{(0)}_{n}(t)}\phi_{n}^{(0)}(x,t) \, , 
\label{eq:SCH-I0}
\end{equation}
is indeed a solution, where $\theta_{n}^{(0)}(t)$ is determined after substituting~\eqref{eq:SCH-I0} in~\eqref{eq:schrH0}. It takes the following expression~\cite{Lew69,Zel19}
\begin{equation}
\theta_{n}^{(0)}(x,t)=-(2n+1)\int^{t}\frac{dt'}{\sigma^{2}(t')}=-\left( n+\frac{1}{2} \right) \arctan\left[\frac{W_{0}}{2}\left(\sqrt{ac-\frac{4}{W_{0}^{2}}}+c\frac{q_{2}(t)}{q_{1}(t)} \right) \right] \, .
\label{eq:CHI0}
\end{equation}
Contrary to the stationary case, the time-dependent complex phase does not represent the time-evolution of the system. 

In summary, to completely determine the solutions of the parametric oscillator, we only need to find two linearly independent solutions of~\eqref{eq:lineal}, provided that $\Omega(t)$ has been already specified. 




\section*{Acknowledgment}
K. Zelaya acknowledges the support from the Mathematical Physics Laboratory of the Centre de Recherches Mat\'ematiques, through a postdoctoral fellowship. He also acknowledges the support of
Consejo Nacional de Ciencia y Tecnolog\'ia (Mexico), grant number A1-S-24569. V. Hussin acknowledges the support of research grants from NSERC of Canada. I. Marquette was supported by Australian Research Council Future Fellowship FT180100099.


\end{document}